\pdfoutput=1
\documentclass[aps,twocolumn,prx,amsfonts,showpacs,superscriptaddress,longbibliography]{revtex4-2}
\usepackage{graphicx}
\usepackage{bm}
\usepackage{hyperref}
\usepackage{amsmath}
\usepackage{amssymb}
\usepackage[table]{xcolor}
\usepackage{array}
\usepackage{multirow}
\usepackage[caption=false]{subfig}
\usepackage{physics}
\usepackage{booktabs}
\usepackage{soul}
\usepackage{amsthm}
\usepackage{bbm}
\usepackage{multirow}
\usepackage[colorinlistoftodos]{todonotes}
\usepackage[scr=boondoxo,frak=boondox,scrscaled=1.05]{mathalfa}
\usepackage{amsmath}

\def\ket#1{|#1\rangle}

\usepackage[normalem]{ulem}

\begin{document}

\title{Non-Abelian quantum Hall states at strong Landau level mixing: \\ finding anyons with first-principles AI}

\author{Pierre-Antoine Graham}
\thanks{These authors contributed equally to this work.}
\affiliation{Department of Physics, Massachusetts Institute of Technology, Cambridge, Massachusetts 02139, USA}
\author{Ahmed Abouelkomsan}
\thanks{These authors contributed equally to this work.}
\affiliation{Department of Physics, Massachusetts Institute of Technology, Cambridge, Massachusetts 02139, USA}
\author{Liang Fu}
\affiliation{Department of Physics, Massachusetts Institute of Technology, Cambridge, Massachusetts 02139, USA}

\begin{abstract}
Non-Abelian anyons, strange particles forming the basis of fault-tolerant topological quantum computing, are predicted to emerge in fractional quantum Hall states at even-denominator filling factors. Identifying which topological order is realized has proven to be a challenging problem that requires an accurate, non-perturbative treatment of realistic Landau level mixing, which has so far been missing from theoretical studies.  %
Here we present the first complete solution of non-Abelian %
quantum Hall problem  that includes \textit{all} electrons and \textit{all} Landau levels, for spin-polarized electrons at filling factor $\nu = 3/2$. We develop a momentum-resolved neural-network variational wave function, optimized solely on the microscopic Hamiltonian, which gives direct access to the topologically degenerate ground states. Without any physics input about the expected outcome, it discovers a degenerate ground-state manifold whose entanglement reveals anyons with quantum dimension $\sqrt{2}$, the hallmark of Ising-type non-Abelian order. By directly computing the overlap with trial wave functions, we further show that Landau level mixing favors the anti-Pfaffian over other candidate states. Our results demonstrate the power of neural-network wave functions in finding and characterizing non-Abelian anyons from first principles, in regimes of strong Landau level mixing, relevant to many recent experiments.
\end{abstract}

\maketitle

\textit{Introduction.}--- Non-Abelian anyons are strange particles that emerge in quantum many-body systems: their quantum state depends on the order in which they are exchanged. Exchanging two of them does not merely multiply the underlying wave function by a phase, as for bosons and fermions, but rotates it within a degenerate manifold of states~\cite{moore1991nonabelions, wen1991non, nayak2008non}. Because these states cannot be distinguished by any local measurement, information stored in them is protected from decoherence, which makes non-Abelian anyons potentially useful for fault-tolerant topological quantum computation~\cite{kitaev2003fault, nayak2008non}. Among the possible platforms, fractional quantum Hall states at even-denominator filling factors are arguably the most promising.

\begin{figure*}[t]
\centering
\includegraphics{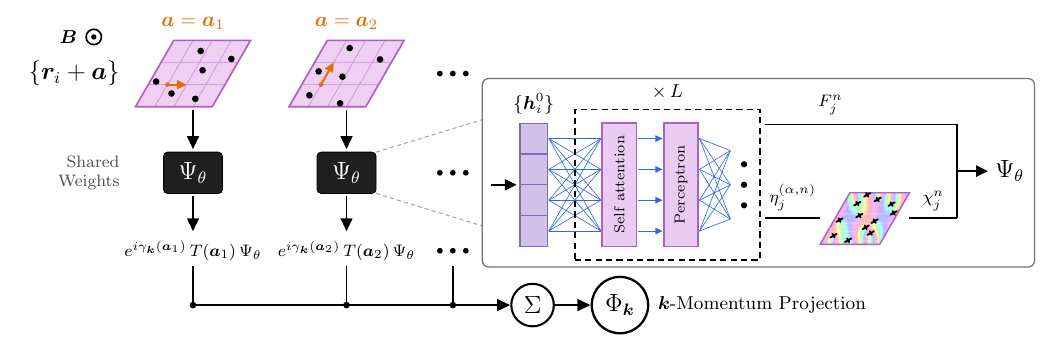}
\caption{
Schematic illustration of the momentum-resolved MagNet ansatz. Electron coordinates are mapped to high-dimensional features $\{\boldsymbol{h}_i\}$ that are periodic on the torus and then passed through $L$ layers combining self-attention and multilayer perceptrons. The network outputs the periodic amplitudes $F^n_j$ and the winding maps $\eta^{(\alpha,n)}_j$, which parametrize the zeros (crosses) of the quasiperiodic factor $\chi^n_j$, shown on the supercell with its phase encoded in color. Together, $F^n_j$ and $\chi^n_j$ form the many-body orbitals of the variational wave function $\Psi_{\theta}$ in Eq.~\eqref{eq:ansatz}. The momentum eigenstate $\Phi_{\mathbf{k}}$ is then obtained by summing the translated copies of $\Psi_{\theta}$ over the $N_\phi$ center-of-mass magnetic translations $T(\mathbf{a})$, as in Eq.~\eqref{eq:kprojection}.}
\label{fig:architecture}
\end{figure*}

The first even-denominator fractional quantum Hall state was observed at $\nu = 5/2$ in GaAs, where the lowest Landau level is completely filled for both spin species and the first excited Landau level is half filled~\cite{willett1987observation}. It was proposed that this state is the Moore--Read Pfaffian~\cite{moore1991nonabelions, greiter1991paired}, a paired state of composite fermions hosting non-Abelian anyons. The Pfaffian, however, is not the only possibility. In the absence of Landau level mixing, the half-filled Landau level has a particle--hole symmetry that maps the Pfaffian onto a distinct topological order, the anti-Pfaffian~\cite{levin2007particle, lee2007particle}. In addition to these two candidates, a third, particle--hole Pfaffian (the PH-Pfaffian), which can be particle-hole symmetric, has been proposed~\cite{son2015composite}. Which of these topological orders is realized has proven remarkably subtle~\cite{ma2024fractional}. Thermal transport measurements are consistent with the PH-Pfaffian~\cite{banerjee2018observation, dutta2022distinguishing, dutta2022isolated, paul2024topological}, in contrast to numerical studies~\cite{morf1998transition, rezayi2000incompressible, feiguin2008density, rezayi2011breaking, zaletel2015infinite, pakrouski2015phase, rezayi2017landau}. This tension has motivated various proposals involving disorder and edge physics~\cite{simon2018interpretation, feldman2018comment, simon2018reply, ma2019partial, simon2020partial, mross2018theory, wang2018topological, lian2018theory, lotric2025majorana}, and more than three decades after its discovery, the nature of the $\nu = 5/2$ state remains an open question.

Beyond the $\nu = 5/2$ state in GaAs, even-denominator states have been recently observed in a growing number of platforms, in which Landau level mixing is often much stronger. In wide and double GaAs quantum wells, a state at $\nu = 1/2$ forms in the lowest Landau level~\cite{suen1992observation, eisenstein1992new, suen1992correlated, suen1994origin}, with an energy gap which reached about 6~K, the largest reported for any even-denominator state~\cite{singh2025fractional}. In GaAs hole systems, fractional quantum Hall states appear at several even-denominator fillings of the lowest Landau level, in the presence of significant Landau level mixing~\cite{wang2023next}. Even-denominator states have also been observed in ZnO~\cite{falson2015even}, in AlAs~\cite{hossain2018unconventional}, in bilayer graphene~\cite{ki2014observation, li2017even, zibrov2017tunable, huang2022valley, assouline2024energy}, and in monolayer WSe$_2$~\cite{shi2020odd, pack2024charge}. In platforms such as ZnO and WSe$_2$, the large effective mass makes the Coulomb energy comparable considerably larger than the cyclotron energy, so that Landau level mixing cannot be treated as a small perturbation.

 Previous studies of the fractional quantum Hall problem have not been able to fully tackle Landau level mixing, which is crucial for resolving distinct topological orders at even-denominator fillings~\cite{antonic2018paired}. Numerical calculations have either truncated the Hilbert space to a single or a few Landau levels~\cite{rezayi2011breaking, zaletel2015infinite}, or treated Landau level mixing perturbatively~\cite{bishara2009effect, simon2013landau, peterson2013more, sodemann2013landau, pakrouski2015phase, rezayi2017landau, herviou2023possible}. These approaches cannot treat the fractional quantum Hall problem at strong Landau level mixing.

In recent years, a new approach based on neural-network wave functions has been developed for solving the fractional quantum Hall problem directly in continuum real space  ~\cite{teng2025solving, qian2025describing, abouelkomsan2026first, perezfadon2025extracting}. Remarkably, without any physics pre-knowledge, the self-attention neural network, optimized purely by energy minimization, achieves very high accuracy beating Landau level projected exact diagonalization \cite{teng2025solving}. Furthermore, the same neural network discovers the transition to electron crystal states at very strong Landau level mixing \cite{abouelkomsan2026first}.

In this work, we present the first solution of the non-Abelian fractional quantum Hall problem in the full Hilbert space, which include \textit{all} electrons in \textit{all} Landau levels. Our neural network solution encompasses the full range of Landau level mixing values from weak to ultra strong. This allows us to study the stability of a non-Abelian state against strong Landau level mixing and determine its topological order, all from first-principles alone and without any physics input about the expected ground state.

\textit{Setup}--- We consider $N$ spin polarized electrons confined to a plane with perpendicular magnetic field $B_0$ and interacting with Coulomb repulsion. The Hamiltonian reads
\begin{align}
    H = \sum_{i=1}^N \frac{[-i \hbar \boldsymbol{\nabla}_i + e \boldsymbol{A}(\mathbf{r}_i)]^2}{2m} + \frac{1}{2}\sum_{i\neq j} \frac{e^2}{4\pi \epsilon|\mathbf{r}_i - \mathbf{r}_j|}  \label{eq:hamiltonian}
\end{align}
where $\mathbf{A} = B_0(\hat{z} \times \mathbf{r})/2$. The problem has two energy scales: the cyclotron energy $\hbar\omega_c = \hbar eB_0/m$, which sets the spacing between Landau levels, and the Coulomb energy $U = e^2/(4\pi\epsilon\ell_B)$, where $\ell_B = \sqrt{\hbar/eB_0}$ is the magnetic length. Their ratio defines the Landau level mixing parameter $\kappa = U/\hbar\omega_c$. We further define the filling factor $\nu = N/N_\phi$, where $N_\phi$ is the number of magnetic flux quanta through the system. Equivalently, $\kappa = \ell_B/a_B \propto B_0^{-1/2}$, where $a_B = 4\pi\epsilon\hbar^2/(me^2)$ is the effective Bohr radius. In terms of the electron density $n_e = \nu/(2\pi\ell_B^2)$, $\kappa = r_s\sqrt{\nu/2}$ with $r_s = 1/(a_B\sqrt{\pi n_e})$, so that $\kappa$ is experimentally tunable with the density and the magnetic field.

We work on a torus spanned by two vectors $\mathbf{L}_1$ and $\mathbf{L}_2$, defining a supercell that encloses an integer number $N_\phi = |\mathbf{L}_1 \times \mathbf{L}_2|/(2\pi\ell_B^2)$ magnetic flux quanta. On this geometry, any wave function $\Psi(\mathbf{r}_1, \ldots, \mathbf{r}_N)$ must satisfy the magnetic boundary conditions $\Psi(\mathbf{r}_1, \ldots, \mathbf{r}_i + \mathbf{L}, \ldots, \mathbf{r}_N) = e^{i \xi_{\mathbf{L}}(\mathbf{r}_i)}\, \Psi(\mathbf{r}_1, \ldots, \mathbf{r}_N)$ for any supercell translation $\mathbf{L} = p\mathbf{L}_1 + q\mathbf{L}_2$ with integers $p$ and $q$, where the position-dependent magnetic phase $\xi_{\mathbf{L}}(\mathbf{r})$ satisfies $\boldsymbol{\nabla}\xi_{\mathbf{L}}(\mathbf{r}) = (e/\hbar)\left[\boldsymbol{A}(\mathbf{r} + \mathbf{L}) - \boldsymbol{A}(\mathbf{r})\right]$. 

Here we build on the MagNet neural network wavefunction introduced in Ref.~\cite{abouelkomsan2026first}, which satisfies these boundary conditions exactly. We consider a variational ansatz of the form
\begin{align}
    \Psi_{\theta}(\mathbf{r}_1, \ldots, \mathbf{r}_N) = \sum_{n=1}^{N_{\rm det}} \det\left(\phi_j^n(\mathbf{r}_i; \{\mathbf{r}_{\neq i}\})\right),
    \label{eq:ansatz}
\end{align}
where $\theta$ denotes the variational parameters and $\phi_j^n(\mathbf{r}_i; \{\mathbf{r}_{\neq i}\})$ is a many-body orbital for particle $i$ that depends on the positions of all particles.  It is symmetric under permutations of the remaining coordinates $\{\mathbf{r}_{\neq i}\}$, so that antisymmetry under particle exchange is realized entirely by the determinant. Importantly, the dependence on the remaining coordinates allows to capture correlation effects beyond mean field \cite{pfau2020ab,von2022self,geier2025self,pescia2024message,smith2024unified}.

In MagNet \cite{abouelkomsan2026first}, each orbital factorizes as
\begin{align}
    \phi_j^n(\mathbf{r}_i; \{\mathbf{r}_{\neq i}\}) = F^n_j(\mathbf{r}_i; \{\mathbf{r}_{\neq i}\}) \, \chi^n_j(\mathbf{r}_i; \{\mathbf{r}_{\neq i}\}),
    \label{eq:orbital}
\end{align}
where $F^n_j$ is periodic on the torus in all particle coordinates and $\chi^n_j$ is quasiperiodic in $\mathbf{r}_i$, carrying the magnetic boundary conditions. The quasiperiodic part is parametrized as
\begin{align}
    \chi^{n}_j(\mathbf{r}_i; \{\mathbf{r}_{\neq i}\}) = \prod_{\alpha = 1}^{N_\phi} f\left(z_i  - \eta^{(\alpha, n)}_j(\mathbf{r}_i; \{\mathbf{r}_{\neq i}\})\right),
    \label{eq:chi}
\end{align}
where $z_i = x_i + i y_i$ is the complex coordinate of particle $i$, $f$ is a fixed quasiperiodic function with unit winding around the supercell boundary, and $\eta^{(\alpha, n)}_j(\mathbf{r}_i; \{\mathbf{r}_{\neq i}\})$ are learnable winding maps, periodic in all particle coordinates, that control the zeros of the orbital while fixing its total winding to $N_\phi$. We refer to Ref.~\cite{abouelkomsan2026first} and the Supplemental Material (SM)~\cite{supplementary} for details.

A hallmark of topologically ordered states, such as fractional quantum Hall states, is topological ground-state degeneracy on surfaces of nonzero genus, such as the torus~\cite{wen1990ground}. In the presence of translation symmetry, the degenerate ground states are further labeled by momentum eigenvalues~\cite{haldanemanybody}, whose pattern provides an additional fingerprint of the topological order.

A generic variational wave function $\Psi_{\theta}$, however, need not carry definite momentum and can mix different sectors. To construct states of definite momentum, we exploit the magnetic translation symmetry of the Hamiltonian in Eq.~\eqref{eq:hamiltonian}. We first choose two generators of single-particle magnetic translations, $t(\mathbf{a}_1)$ and $t(\mathbf{a}_2)$, where the vectors $\mathbf{a}_1$ and $\mathbf{a}_2$ are chosen to enclose a single flux quantum, $|\mathbf{a}_1 \times \mathbf{a}_2| = 2\pi\ell_B^2$, so that the generators commute. Together they define $N_\phi$ distinct translations $t(\mathbf{a})$, with $\mathbf{a} = m\mathbf{a}_1 + n\mathbf{a}_2$ for integers $m$ and $n$, that span the supercell. From these we build the center-of-mass (COM) translations $T(\mathbf{a}) = \prod_{i=1}^{N} t_i(\mathbf{a})$, which displace all particles simultaneously by $\mathbf{a}$. They commute with each other and with $H$, and their simultaneous eigenstates are labeled by a center-of-mass momentum $\mathbf{k}$ taking $N_\phi$ distinct values.

We construct COM momentum eigenstates from $\Psi_{\theta}$ by summing its translated copies over the $N_\phi$ translations $T(\mathbf{a})$:
\begin{align}
    \Phi_{\mathbf{k}}(\mathbf{r}_1, \ldots, \mathbf{r}_N)
    = \frac{1}{N_\phi} \sum_{\mathbf{a}}
    e^{i\gamma_{\mathbf{k}}(\mathbf{a})}\, T(\mathbf{a})\,
        \Psi_{\theta}(\mathbf{r}_1, \ldots, \mathbf{r}_N),
    \label{eq:kprojection}
\end{align}
where $\gamma_{\mathbf{k}}(\mathbf{a})$ are appropriately chosen phases such that $\Phi_{\mathbf{k}}$ carries momentum $\mathbf{k}$~\cite{supplementary,abouelkomsan2025topological}.

We use a neural network based on self-attention~\cite{vaswani2017attention, von2022self,  geier2025self, li2025attention,abouelkomsan2026first} to represent the many-body orbitals $\phi^n_j$ in Eq.~\eqref{eq:ansatz}. In Fig.~\ref{fig:architecture}, we present a schematic overview of the architecture. In neural-network variational Monte Carlo (NN-VMC), the network parameters $\theta$ are optimized to minimize the variational energy $E_{\mathbf{k}}(\theta) = \langle \Phi_{\mathbf{k}} | H | \Phi_{\mathbf{k}} \rangle / \langle \Phi_{\mathbf{k}} | \Phi_{\mathbf{k}} \rangle$, which, together with its gradients, is estimated by Monte Carlo sampling.

\textit{Non-Abelian FQH at $\nu = 3/2$.}--- Next, we study filling factor $\nu = 3/2$. In the absence of Landau level mixing, such a filling corresponds, in our fully spin-polarized setting, to a completely filled lowest Landau level (LLL) and a half-filled first excited Landau level (1LL). We optimize $\Phi_{\mathbf{k}}$ independently in each of the $N_\phi$ momentum sectors, which yields a momentum-resolved low-energy spectrum that can be compared directly with exact diagonalization (ED). For the latter, we diagonalize Eq.~\eqref{eq:hamiltonian} with the Hilbert space restricted to the LLL and 1LL.

\begin{figure}[t]
\centering
\includegraphics[width=\columnwidth]{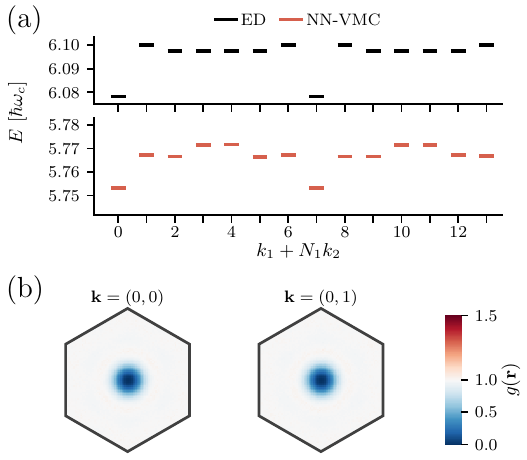}
\caption{
(a) Lowest energy in each momentum sector $\mathbf{k} = (k_1, k_2)$, labeled by $k_1 + N_1 k_2$, for $N = 21$ electrons and $\kappa = 1.0$, obtained from NN-VMC and from ED restricted to the LLL and 1LL. We use a hexagonal supercell spanned by $\mathbf{L}_1 = N_1 \mathbf{a}_1$ and $\mathbf{L}_2 = N_2 \mathbf{a}_2$ with $N_1 = 7$ and $N_2 = 2$, so that $k_i = 0, \ldots, N_i - 1$. The NN-VMC energies lie well below the ED energies in all sectors, and both show a twofold degenerate ground state at $\mathbf{k} = (0,0)$ and $(0,1)$. (b) Pair correlation function $g(\mathbf{r})$ of the two NN-VMC ground states.}
\label{fig:spectrum}
\end{figure}

\begin{figure}[t]
\centering
\includegraphics[width=\columnwidth]{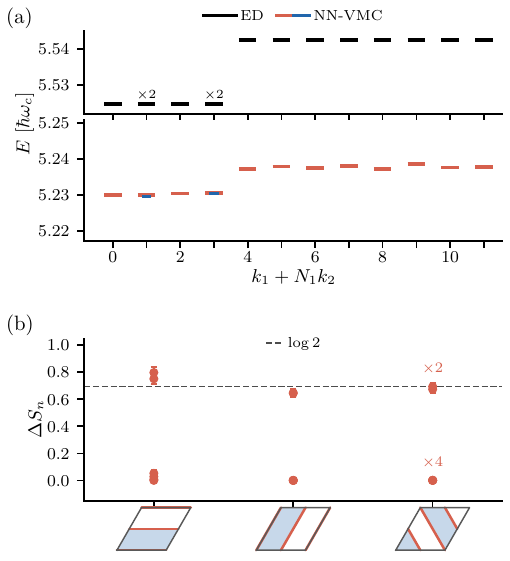}
\caption{
(a) Lowest energies in each momentum sector $\mathbf{k} = (k_1, k_2)$, labeled by $k_1 + N_1 k_2$, for $N = 18$ electrons and $\kappa = 1.0$, obtained from NN-VMC and from ED restricted to the LLL and 1LL. We use a hexagonal supercell spanned by $\mathbf{L}_1 = N_1 \mathbf{a}_1$ and $\mathbf{L}_2 = N_2 \mathbf{a}_2$ with $N_1 = 4$ and $N_2 = 3$. Numbers indicate twofold degenerate ED levels within a sector. The blue labeled states are orthogonal ground states obtained with NN-VMC. (b) Differences $\Delta S_n$ between the second R\'enyi entropies of the six MESs and the smallest one, for three inequivalent noncontractible cuts of the torus (bottom). For each cut, four MESs have $\Delta S_n \approx 0$, consistent with Abelian sectors, while the remaining two lie close to $\log 2$ (dashed line), as expected for $d_\sigma = \sqrt{2}$.}
\label{fig:entropy}
\end{figure}

Figure~\ref{fig:spectrum} shows the resulting spectra for $N = 21$ electrons and $N_\phi = 14$ at Landau level mixing $\kappa = 1.0$. Remarkably, in every momentum sector the NN-VMC energies are significantly lower than the ED energies. In addition, the spectrum features a twofold degenerate ground state, separated by a gap from the excited states. This is precisely the degeneracy expected for a Moore--Read-type non-Abelian state with an odd number of electrons in the half-filled active Landau level. On the torus, the Moore--Read state is sixfold degenerate for an even number of particles but only twofold degenerate for an odd number~\cite{read2000paired, oshikawa2007topological, ardonne2008degeneracy}. Consistently, for $N = 18$, where the 1LL hosts an even number of electrons, we find a sixfold degenerate ground-state manifold [Fig.~\ref{fig:entropy}(a)].

Moreover, the pair correlation function $g(\mathbf{r}) = (2\pi\ell_B^2/\nu N) \langle \sum_{i \neq j} \delta(\mathbf{r} - \mathbf{r}_i + \mathbf{r}_j) \rangle$ of the two ground states, shown in Fig.~\ref{fig:spectrum}(b), exhibits a correlation hole at short distances and is otherwise featureless, confirming the expected liquid nature of the ground states.

\textit{Minimally entangled states and quantum dimensions.}---
Having established ground-state degeneracy consistent with the non-Abelian Moore--Read state, we now turn to probe its anyon content directly. In addition to topological degeneracy, topological order exhibits long-range entanglement in the ground states, which cannot be detected by any local observable~\cite{wen1990topological, wen2004quantum}. In particular, the quantum dimensions of the anyons are imprinted in the topological entanglement entropy~\cite{kitaev2006topological, levin2006detecting, dong2008topological}. On the torus, it was shown that this information can be extracted from the minimally entangled states (MESs), which are linear superpositions of the underlying degenerate ground states that minimize the entanglement entropy for a noncontractible bipartition~\cite{zhang2012quasiparticle}. Each MES carries a definite quasiparticle type $a$ threading the cut, and its R\'enyi entanglement entropy $S_n = (1-n)^{-1}\log\operatorname{Tr}\rho_A^n$, with $\rho_A$ the reduced density matrix obtained after tracing out one part of the system, takes the form~\cite{flammia2009topological, zhang2012quasiparticle}
\begin{equation}
    S_n^{(a)} = \alpha_n L - 2\log\left(\frac{\mathcal{D}}{d_a}\right) + \cdots,
    \label{eq:MESentropy}
\end{equation}
where $L$ is the total length of the boundary between the two parts, $\alpha_n$ is a nonuniversal coefficient, $d_a$ is the quantum dimension of anyon $a$, and $\mathcal{D} = \sqrt{\sum_a d_a^2}$ is the total quantum dimension. Differences between the entropies of MESs therefore cancel the boundary contribution and directly measure ratios of quantum dimensions, $S_n^{(a)} - S_n^{(b)} = 2\log(d_a/d_b)$~\cite{zhu2014identifying}.

The quasiparticles of the fermionic Moore--Read state combine the Ising anyons $\{1, \psi, \sigma\}$, where $\psi$ is a neutral Majorana fermion and $\sigma$ the non-Abelian Ising anyon, with a $U(1)$ charge sector~\cite{moore1991nonabelions, read2000paired, hansson2017quantum}. Its six ground states on the torus correspond to six quasiparticle types: four Abelian ones, built from $1$ and $\psi$ with charge $0$ or $e/2$, which have $d_a = 1$, and two non-Abelian ones, containing $\sigma$ with charge $e/4$ or $3e/4$, which have $d_\sigma = \sqrt{2}$. The total quantum dimension is thus $\mathcal{D} = 2\sqrt{2}$. The four Abelian MESs are therefore expected to have equal entropies, while the two non-Abelian MESs should have larger entropies by
\begin{equation}
    \Delta S_n = 2\log\sqrt{2} = \log 2.
    \label{eq:entropy_spectrum}
\end{equation}

To access all six topological sectors, we perform this analysis for $N = 18$, where the 1LL hosts an even number of electrons and the ground-state manifold is sixfold degenerate. In this case, some of the six ground states share the same momentum sector [Fig.~\ref{fig:entropy}(a)]. These degenerate states are distinguished by their eigenvalues under a finer center-of-mass magnetic translation, and we resolve them by an additional projection onto these eigenvalues~\cite{supplementary}. As an independent check, we also obtain the degenerate partners by including in the loss function a penalty on the overlap with previously converged states, which yields the same results~\cite{supplementary}. We then construct the MESs as normalized superpositions
\begin{equation}
    \ket{\Omega_a} = \sum_{\alpha=1}^{6} c^{(a)}_\alpha \ket{\Phi_\alpha}, \qquad a = 1, \ldots, 6,
    \label{eq:superposition}
\end{equation}
of the six NN-VMC ground states $\ket{\Phi_\alpha}$, and evaluate the second R\'enyi entropy $S_2$ for a noncontractible cut of the torus using the replica (swap) estimator~\cite{hastings2010measuring, tubman2012renyi, perezfadon2025extracting}. The coefficients $c^{(a)}_\alpha$ are then optimized to minimize $S_2$~\cite{perezfadon2025extracting}.

As shown in Fig.~\ref{fig:entropy}(b), the entropies of the six MESs are arranged exactly as predicted for the Moore--Read state: four MESs have nearly equal entropies, identifying them with the Abelian sectors, while the remaining two are larger by an amount consistent with $\log 2$, which fixes their quantum dimension to $d_\sigma = \sqrt{2}$. This structure of the entropies provides direct evidence of the Ising-type non-Abelian topological order of the Moore--Read state.

\textit{Landau-level mixing.}--- So far, we have identified the topological order of the ground state at $\kappa = 1$. We now study the evolution of the many-body ground states with Landau level mixing. Its importance is already visible in Fig.~\ref{fig:spectrum}(a), where the NN-VMC energies lie well below the ED energies obtained within the LLL and 1LL, showing that the higher Landau levels substantially lower the energy.

\begin{figure}[t]
\centering
\includegraphics[width=\columnwidth]{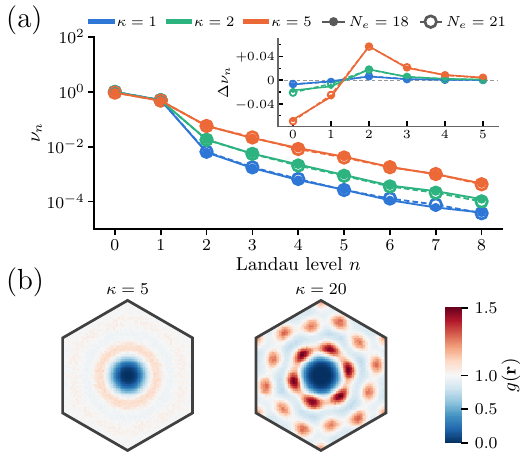}
\caption{(a) Occupation $\nu_n$ of Landau level $n$, on a logarithmic scale, for $N = 18$ and $N = 21$ electrons at $\kappa = 1, 2$ and $5$. Inset: deviation $\Delta\nu_n$ from the unmixed occupations $\nu_0 = 1$, $\nu_1 = 1/2$, and $\nu_{n \geq 2} = 0$. Both the LLL and the 1LL are depleted, with the weight transferred to higher levels. (b) Pair correlation function computed at $\kappa = 5$ and $20$ (see \cite{supplementary} for more details). }
\label{fig:LLmixing}
\end{figure}

To quantify the effect of Landau level mixing on the non-Abelian state, we extract the occupation $\nu_n$ of the $n$th Landau level from the one-body density matrix of the converged states~\cite{supplementary}, with $\sum_n \nu_n = \nu = 3/2$. Figure~\ref{fig:LLmixing}(a) shows $\nu_n$ for $\kappa = 1$, $2$, and $5$, for both $N = 18$ and $N = 21$. Without Landau level mixing, $\nu_0 = 1$, $\nu_1 = 1/2$, and all higher levels are empty. With Landau level mixing, the higher levels acquire a sizable occupation, which decays with $n$ and grows with $\kappa$. The occupations for the two system sizes nearly coincide, indicating that finite-size effects are small.

Moreover, the inset of Fig.~\ref{fig:LLmixing} shows the deviation $\Delta\nu_n$ of each Landau level occupation from its value in the absence of Landau level mixing. As expected, the LLL is depleted. Less obviously and more interestingly, the 1LL is depleted as well, even though its occupation could in principle increase due to electrons scattering from the LLL. The weight lost from both levels is instead transferred to $n \geq 2$. We find this feature to persist for various values of $\kappa$ for both system sizes.

Beyond redistributing the electrons among the Landau levels, strong Landau level mixing eventually changes the nature of the ground state \cite{abouelkomsan2026first, reddy2026quantum}. Figure~\ref{fig:LLmixing}(b) contrasts the pair correlation function at $\kappa = 5$ and $\kappa = 20$. At $\kappa = 5$, $g(\mathbf{r})$ shows only a correlation hole and is otherwise featureless, as expected for a liquid. At $\kappa = 20$, it instead develops pronounced peaks arranged on a triangular lattice, consistent with a Wigner crystal.

\begin{figure}[t]
\centering
\includegraphics[width=\columnwidth]{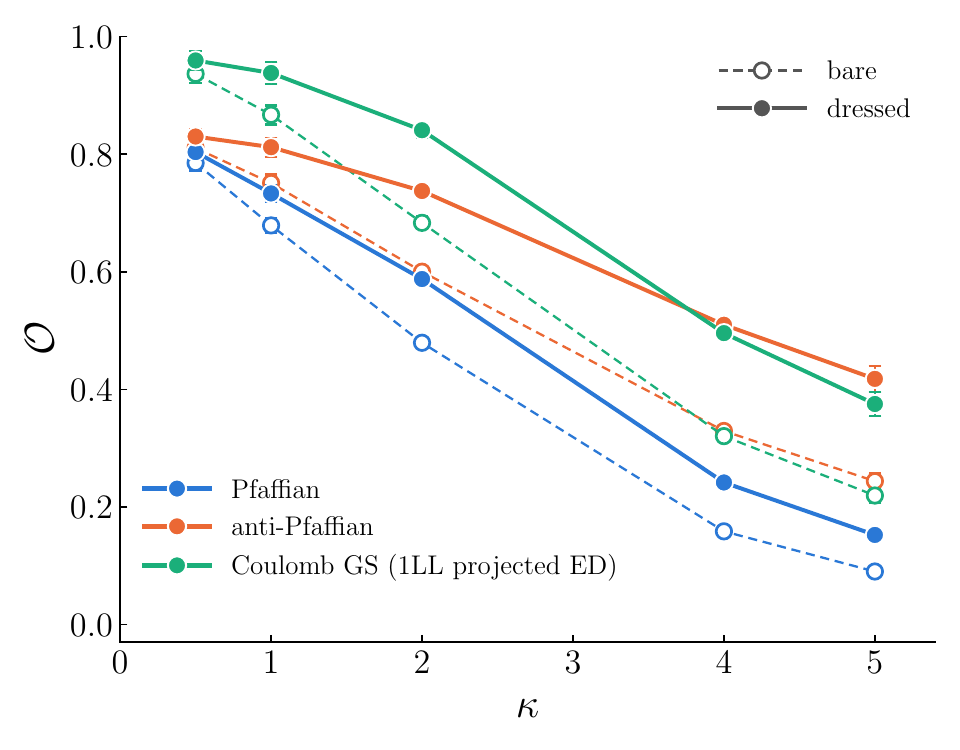}
\caption{
Overlap $\mathcal{O}$ between the NN-VMC ground-state manifold and the Moore--Read Pfaffian, the anti-Pfaffian, and the exact Coulomb ground state of the 1LL-projected Hamiltonian, as a function of Landau level mixing $\kappa$, for $N = 18$ electrons. Open symbols with dashed lines show overlaps with the bare trial states, defined within the LLL and 1LL, and filled symbols with solid lines show overlaps with dressed trial states, which include a cloud of virtual inter-Landau-level excitations~\cite{supplementary}. In both cases, the overlap with the anti-Pfaffian exceeds that with the Pfaffian at all $\kappa$.
}
\label{fig:pfapf}
\end{figure}
\textit{Candidate topological order.}---
In the absence of Landau level mixing ($\kappa \to 0$), the electrons in the LLL at $\nu = 3/2$ are frozen and the problem reduces to a half-filled 1LL. In this limit, the Hamiltonian has an exact particle--hole symmetry about half filling of the 1LL~\cite{levin2007particle, lee2007particle}. However, this symmetry is broken by Landau level mixing at nonzero $\kappa$~\cite{bishara2009effect, simon2013landau, peterson2013more, sodemann2013landau, antonic2018paired}.

Two leading candidates for the non-Abelian topological order are the Moore--Read Pfaffian and its particle--hole conjugate, the anti-Pfaffian~\cite{moore1991nonabelions, levin2007particle, lee2007particle}. They describe distinct topological phases that differ in their edge structure: all edge modes of the Pfaffian propagate in the same direction, whereas the anti-Pfaffian hosts counter-propagating neutral modes~\cite{levin2007particle, lee2007particle}. Being related by particle--hole conjugation, the two states are degenerate only when this symmetry is exact, that is, in the absence of Landau level mixing. A third proposal, the particle--hole symmetric Pfaffian (PH-Pfaffian)~\cite{son2015composite}, is instead invariant under particle--hole conjugation. To study the competition between these candidates as Landau level mixing is turned on, we compare the manifold of optimized NN-VMC ground states with the corresponding trial manifolds. We define the overlap between the NN-VMC and trial manifolds as $\mathcal{O} = d^{-1} \sum_{\alpha, \beta = 1}^{d} | \langle \Phi_\alpha | \mathrm{X}_\beta \rangle |^2$, where $\{\ket{\Phi_\alpha}\}$ and $\{\ket{\mathrm{X}_\beta}\}$ are orthonormal bases of the $d$-dimensional NN-VMC and trial ground-state manifolds, respectively, with $\mathrm{X}$ labeling the trial state. Since the trial wave functions are usually defined in the LLL, we promote them to the 1LL and add a background of a frozen, completely filled LLL~\cite{supplementary}. Because the NN-VMC states also have weight outside the LLL and 1LL, we further consider dressed trial states, in which a cloud of virtual inter-Landau-level excitations is attached to each trial state through a Schrieffer--Wolff rotation to first order in Landau level mixing, whose strength is optimized at each $\kappa$~\cite{supplementary}. 

Figure~\ref{fig:pfapf} shows $\mathcal{O}$ for $N = 18$, where $d = 6$, as a function of $\kappa$ up to $\kappa = 5$ for the Pfaffian, the anti-Pfaffian and, as a particle--hole symmetric reference, the exact Coulomb ground state obtained by ED within the 1LL, both for bare and dressed trial states. The overlap with the anti-Pfaffian is consistently larger than with the Pfaffian, showing that Landau level mixing favors the anti-Pfaffian. As $\kappa$ decreases, the two overlaps approach each other, as expected from the restoration of particle--hole symmetry in the limit $\kappa \to 0$. As is evident, dressing the trial states increases all overlaps, highlighting the importance of Landau level mixing, which the NN-VMC states take into account.

Generally, the overlaps decrease with increasing $\kappa$, reflecting the growing weight of the NN-VMC states outside the LLL and 1LL [Fig.~\ref{fig:LLmixing}(a)]. Remarkably, at large $\kappa$, the overlap with the anti-Pfaffian becomes comparable to, and even exceeds, that with the 1LL-projected Coulomb ground state.

We also computed the overlap with several forms of the PH-Pfaffian~\cite{supplementary}. In contrast to the Pfaffian and the anti-Pfaffian, which are the zero modes of parent Hamiltonians, it is not clear what a good trial wave function for the PH-Pfaffian is~\cite{mishmash2018numerical, yang2023second}. The forms we considered either have a negligible overlap with the NN-VMC ground states or break particle--hole symmetry and are closely related to the anti-Pfaffian~\cite{yang2023second, zhao2023composite, supplementary}.

\textit{Discussion.}---
In this work, we have studied the non-Abelian fractional quantum Hall state of spin-polarized electrons at $\nu = 3/2$, taking all Landau levels into account non-perturbatively. Starting from the microscopic Hamiltonian of interacting electrons in a magnetic field, and without any input about the expected ground state, our approach captures the topological order, the Landau level mixing, and the competition between candidate states within a single framework. We find that Landau level mixing favors the anti-Pfaffian topological order over other candidate states.

Our approach differs from neural-network methods that build physical input into the variational wave function states~\cite{nazaryan2025artificial,gattu2025dressing,badalyan2026neural}. Here, our variational wave function is initialized randomly and optimized solely by minimizing the energy, so that none of the competing topological orders is favored from the start. This is essential for determining which of them is realized.

Our results are directly relevant to recent experiments. An even-denominator state at $\nu = 3/2$ has been observed in monolayer WSe$_2$, both in compressibility~\cite{shi2020odd} and in transport~\cite{pack2024charge} measurements. In this material, spin--valley locking together with a large spin splitting makes the low-lying Landau levels fully spin and valley polarized, so that the state at $\nu = 3/2$ forms in a half-filled 1LL on top of a completely filled LLL, as in the setting studied here. An even-denominator state at $\nu = 3/2$ has also been reported in ZnO~\cite{falson2015even}, where it appears when the partially filled level has the orbital character of the 1LL. In both materials, the large effective mass makes Landau level mixing substantial, which calls for a non-perturbative treatment such as ours.

A natural extension of this work is the $\nu = 5/2$ state in GaAs~\cite{willett1987observation}. There, Landau level mixing differs from the fully spin-polarized case at $\nu = 3/2$ studied here, since it also involves electrons of the opposite spin. Extending our approach to spinful electrons would allow Landau level mixing at $\nu = 5/2$ to be treated on the same footing. Including disorder~\cite{zhu2026crystallization} would further allow studying its effect on the competing topological orders.

Our approach is equally suited to even-denominator states in the lowest Landau level, which have been observed in wide quantum wells and in GaAs hole systems~\cite{singh2025fractional, wang2023next}. There, Landau level mixing has been proposed to induce pairing of composite fermions~\cite{zhao2023composite}. Since our approach treats Landau level mixing non-perturbatively, it can address the nature of these states directly.

\textit{Acknowledgements---} We are grateful to Mansour Shayegan, Cory Dean for insightful discussions. This work was supported by a Simons Investigator Award from the Simons Foundation.
We acknowledge MIT SuperCloud
and the Lincoln Laboratory Supercomputing Center, IAIFI (the NSF Cooperative Agreement PHY-2019786) and MIT Engaging cluster the MIT Office of Research Computing and Data for providing high performance computing resources that have contributed to the research results reported within this paper. We acknowledge the MIT Generative AI Impact Consortium (MGAIC) for providing supporting resources. We acknowledge the support of the Natural Sciences and Engineering Research Council of Canada (NSERC), funding reference number CGRS D-610169-2026.

\textit{Note added.}--- A recent preprint~\cite{ma2026exploring} reports related results.
\bibliography{references}

\onecolumngrid
\newpage
\makeatletter

\begin{center}
\textbf{\large Supplementary materials for:\\ ``{Non-Abelian quantum Hall states at strong Landau level mixing: \\ finding anyons with first-principles AI}''}
\\[10pt]
Pierre-Antoine Graham$^{1*}$, Ahmed Abouelkomsan$^{1*}$, Liang Fu$^{1}$ \\
\textit{$^1$Department of Physics, Massachusetts Institute of Technology, Cambridge, MA-02139, USA}\\
\end{center}
$^*$These authors contributed equally to this work.

\vspace{20pt}

\setcounter{figure}{0}
\setcounter{section}{0}
\setcounter{equation}{0}

\renewcommand{\thefigure}{S\@arabic\c@figure}
\renewcommand{\theequation}{S\@arabic\c@equation}
\renewcommand{\theHequation}{S\@arabic\c@equation}
\renewcommand{\theHfigure}{S\@arabic\c@figure}
\makeatother

\section{Magnetic boundary conditions and momentum projection}\label{sec:bc}

In this section, we describe how the magnetic boundary conditions are imposed on the torus and how we construct states of definite center-of-mass momentum. Two lattices play a role. The supercell, spanned by $\mathbf{L}_1$ and $\mathbf{L}_2$, defines the torus and encloses $N_\phi$ flux quanta. The unit cell, spanned by $\mathbf{a}_1$ and $\mathbf{a}_2$, encloses a single flux quantum, and $N_\phi$ unit cells tile the supercell.

\subsection{Magnetic translations}
In a magnetic field, ordinary translations do not commute with the Hamiltonian, because the vector potential is not translation invariant. They are replaced by magnetic translations, which combine a translation with a gauge transformation. In the symmetric gauge $\mathbf{A} = B_0(\hat{z}\times\mathbf{r})/2$ of the main text, the single-particle magnetic translation by a vector $\mathbf{a}$ acts as
\begin{equation}
    t(\mathbf{a})\,\psi(\mathbf{r}) = e^{-\frac{i}{2\ell_B^{2}}\hat{z}\cdot(\mathbf{a}\times\mathbf{r})}\,\psi(\mathbf{r}+\mathbf{a})
    \label{eq:magtrans}
\end{equation}
and commutes with the kinetic energy. Two magnetic translations compose up to an Aharonov--Bohm phase, $t(\mathbf{a})\,t(\mathbf{b}) = e^{\frac{i}{2\ell_B^{2}}\hat{z}\cdot(\mathbf{a}\times\mathbf{b})}\,t(\mathbf{a}+\mathbf{b})$, so that $t(\mathbf{a})\,t(\mathbf{b}) = e^{\frac{i}{\ell_B^{2}}\hat{z}\cdot(\mathbf{a}\times\mathbf{b})}\,t(\mathbf{b})\,t(\mathbf{a})$. They therefore commute if and only if the parallelogram spanned by $\mathbf{a}$ and $\mathbf{b}$ encloses an integer number of flux quanta.

\subsection{Supercell and boundary conditions}
Since the supercell encloses an integer number $N_\phi$ of flux quanta, translations by any two supercell vectors commute and can be diagonalized simultaneously. The magnetic boundary conditions fix their eigenvalues for every electron. We use the convention of Ref.~\cite{wang2019lattice},
\begin{equation}
    t_i(\mathbf{L})\,\Psi = \eta_{\mathbf{L}}^{N_\phi}\,\Psi, \qquad
    \eta_{\mathbf{L}} = \begin{cases} 1 & \text{if } \mathbf{L}/2 \text{ is a supercell vector,} \\ -1 & \text{otherwise,} \end{cases}
    \label{eq:bcop}
\end{equation}
for every supercell vector $\mathbf{L} = p\mathbf{L}_1 + q\mathbf{L}_2$, without additional twist phases. The sign $\eta_{\mathbf{L}} = (-1)^{pq+p+q}$ makes the conditions for different supercell vectors consistent with the composition rule. Written out with Eq.~\eqref{eq:magtrans}, Eq.~\eqref{eq:bcop} gives the boundary conditions of the main text,
\begin{equation}
    \Psi(\mathbf{r}_1, \ldots, \mathbf{r}_i + \mathbf{L}, \ldots, \mathbf{r}_N) = e^{i\xi_{\mathbf{L}}(\mathbf{r}_i)}\,\Psi(\mathbf{r}_1, \ldots, \mathbf{r}_N), \qquad \xi_{\mathbf{L}}(\mathbf{r}) = \frac{\hat{z}\cdot(\mathbf{L}\times\mathbf{r})}{2\ell_B^{2}} + \pi N_\phi (pq + p + q).
    \label{eq:bc}
\end{equation}
The ansatz of Eqs.~\eqref{eq:ansatz}--\eqref{eq:chi} of the main text satisfies Eq.~\eqref{eq:bc} exactly for every electron, since the amplitudes $F^n_j$ and the winding maps $\eta^{(\alpha,n)}_j$ are periodic on the supercell and the factor $\chi^n_j$ carries the magnetic phase~\cite{abouelkomsan2026first}.

\subsection{Unit cell and single-particle generators}
The Hamiltonian of Eq.~\eqref{eq:hamiltonian} is invariant under magnetic translations of all electrons. As in the main text, we use this symmetry to label states by a center-of-mass momentum. On the torus, the topologically degenerate ground states are then further labeled by their momenta, whose pattern provides an additional fingerprint of the topological order~\cite{haldanemanybody}.

Not every translation is compatible with the boundary conditions. A translation $t(\mathbf{d})$ maps states obeying Eq.~\eqref{eq:bcop} onto states obeying it only if it commutes with $t(\mathbf{L}_1)$ and $t(\mathbf{L}_2)$, that is, if $\mathbf{d}$ spans an integer number of flux quanta with both $\mathbf{L}_1$ and $\mathbf{L}_2$. The smallest such translations are by $\mathbf{L}_1/N_\phi$ and $\mathbf{L}_2/N_\phi$, which span the fine lattice $\Lambda_\phi = \{(m\mathbf{L}_1 + n\mathbf{L}_2)/N_\phi\}$. These two translations do not commute, since the cell they span encloses only $1/N_\phi$ of a flux quantum. To define a momentum, we therefore choose a set of commuting translations on $\Lambda_\phi$. We divide the supercell into $N_\phi = N_1 N_2$ unit cells spanned by $\mathbf{a}_1 = \mathbf{L}_1/N_1$ and $\mathbf{a}_2 = \mathbf{L}_2/N_2$. Each unit cell encloses a single flux quantum, $|\mathbf{a}_1\times\mathbf{a}_2| = 2\pi\ell_B^{2}$, so that $t(\mathbf{a}_1)$ and $t(\mathbf{a}_2)$ commute. Taking them as generators, we obtain $N_\phi$ distinct translations,
\begin{equation}
    t(\mathbf{a}) = t(\mathbf{a}_1)^m\, t(\mathbf{a}_2)^n, \qquad \mathbf{a} = m\mathbf{a}_1 + n\mathbf{a}_2, \quad m = 0, \ldots, N_1 - 1, \quad n = 0, \ldots, N_2 - 1,
    \label{eq:generators}
\end{equation}
since $t(\mathbf{a}_1)^{N_1} = t(\mathbf{L}_1)$ and $t(\mathbf{a}_2)^{N_2} = t(\mathbf{L}_2)$ are fixed by Eq.~\eqref{eq:bcop}. The remaining translations on $\Lambda_\phi$ provide additional quantum numbers, which we discuss in Sec.~\ref{sec:penalty}.

\subsection{Single-particle eigenstates}\label{sec:orbitals}
The generators $t(\mathbf{a}_1)$ and $t(\mathbf{a}_2)$ commute with each other and with the single-particle kinetic energy $\hbar\omega_c\,(a^{\dagger}a + \tfrac12)$, where $a^{\dagger} = (\bar z/2\ell_B - 2\ell_B\,\partial_{z})/\sqrt{2}$ and $z = x + iy$. The single-particle eigenstates can therefore be labeled by a Landau level $n$ and a momentum $\mathbf{q}$. Writing $g(\mathbf{r}) = (2\pi\ell_B^{2})^{-1/2}\,e^{-|z|^{2}/4\ell_B^{2}}$ for the lowest-Landau-level Gaussian centered at the origin, the orbital of momentum $\mathbf{q}$ in the lowest Landau level is the sum of its translated copies over the whole lattice spanned by $\mathbf{a}_1$ and $\mathbf{a}_2$, including the cells of the neighboring supercells~\cite{fremling2013coherent},
\begin{equation}
    \psi_{\mathbf{q}}(\mathbf{r}) = \mathcal{N}_{\mathbf{q}} \sum_{\mathbf{a}} e^{-i\mathbf{q}\cdot\mathbf{a}}\; t(\mathbf{a})\, g(\mathbf{r}),
    \label{eq:lllmomentum}
\end{equation}
with $t(\mathbf{a})$ defined as in Eq.~\eqref{eq:generators} for every lattice vector $\mathbf{a}$, and a normalization constant $\mathcal{N}_{\mathbf{q}}$. On the torus, $\psi_{\mathbf{q}}$ obeys the boundary conditions of Eq.~\eqref{eq:bcop}, which allows $N_\phi$ values of $\mathbf{q}$. The higher Landau levels follow by raising, $\varphi_{n,\mathbf{q}} = (a^{\dagger})^{n}\psi_{\mathbf{q}}/\sqrt{n!}$, since $a^{\dagger}$ commutes with every $t(\mathbf{a})$.

\subsection{Center-of-mass momentum}
We define the center-of-mass translations as $T(\mathbf{a}) = \prod_{i=1}^{N} t_i(\mathbf{a})$. They commute with the Hamiltonian, since the Coulomb interaction depends only on the relative positions of the electrons, and with one another, since the generators do. By Eq.~\eqref{eq:bcop}, $T(\mathbf{a}_j)^{N_j} = \prod_i t_i(\mathbf{L}_j) = (-1)^{N N_\phi}$, so that the joint eigenvalues of $T(\mathbf{a}_1)$ and $T(\mathbf{a}_2)$ are
\begin{equation}
    \lambda_j(\mathbf{k}) = e^{2\pi i (k_j + \delta)/N_j}, \qquad k_j = 0, \ldots, N_j - 1,
    \label{eq:momenta}
\end{equation}
with $\delta = 0$ for even $N N_\phi$ and $\delta = 1/2$ for odd $N N_\phi$. They define a two-dimensional center-of-mass momentum $\mathbf{k} = (k_1, k_2)$ that takes $N_\phi$ distinct values. In Figs.~\ref{fig:spectrum} and~\ref{fig:entropy}, the momentum sectors are labeled by $k_1 + N_1 k_2$.

\subsection{Momentum projection}
A variational wave function $\Psi_\theta$ of the form of Eq.~\eqref{eq:ansatz} of the main text obeys the supercell boundary conditions but is in general a superposition of several momentum sectors. We isolate its component in the sector $\mathbf{k}$ with the projector of Eq.~\eqref{eq:kprojection} of the main text,
\begin{equation}
    \Phi_{\mathbf{k}}(\mathbf{r}_1, \ldots, \mathbf{r}_N) = \frac{1}{N_\phi}\sum_{\mathbf{a}} e^{i\gamma_{\mathbf{k}}(\mathbf{a})}\, T(\mathbf{a})\,\Psi_{\theta}(\mathbf{r}_1, \ldots, \mathbf{r}_N),
    \label{eq:momproj}
\end{equation}
where the sum runs over the $N_\phi$ translations $\mathbf{a} = m\mathbf{a}_1 + n\mathbf{a}_2$ of Eq.~\eqref{eq:generators} and
\begin{equation}
    \gamma_{\mathbf{k}}(\mathbf{a}) = -2\pi\left[\frac{m (k_1 + \delta)}{N_1} + \frac{n (k_2 + \delta)}{N_2}\right].
    \label{eq:gammak}
\end{equation}
Since $e^{i\gamma_{\mathbf{k}}(\mathbf{a})} = \lambda_1(\mathbf{k})^{-m}\,\lambda_2(\mathbf{k})^{-n}$ and $T(\mathbf{a}_j)^{N_j} = \lambda_j(\mathbf{k})^{N_j}$, the projected state is an eigenstate of both generators, $T(\mathbf{a}_j)\,\Phi_{\mathbf{k}} = \lambda_j(\mathbf{k})\,\Phi_{\mathbf{k}}$.

\section{Neural network architecture}\label{sec:architecture}

We use a self-attention neural network to represent  $\Psi_{\theta}$. The input of this network consists of $N$ single-particle streams. In stream $i = 1, \cdots, N$, the coordinates $\mathbf{r}_i$ of electron $i$ enter as a periodized feature vector implementing the periodic boundary conditions for $F_j^n$ and $\eta_j^{(\alpha, n)}$. These features $\mathbf{f}_i$ and their embedding $\mathbf{h}^0_i$ in a $d_L$-dimensional space are given by 
\begin{equation}
    \mathbf{f}_i = \left(\cos \mathbf{G}_1\!\cdot\!\mathbf{r}_i,\;
                         \cos \mathbf{G}_2\!\cdot\!\mathbf{r}_i,\;
                         \sin \mathbf{G}_1\!\cdot\!\mathbf{r}_i,\;
                         \sin \mathbf{G}_2\!\cdot\!\mathbf{r}_i\right)^{\!T},
    \qquad
    \boldsymbol{h}^0_i = W^0 \mathbf{f}_i .
    \label{eq:features}
\end{equation}
where $\mathbf{G}_j$ are the supercell
reciprocal vectors. The particle streams $\{\boldsymbol{h}^0_i\}$ are then propagated through $L$ identical layers, each of which first lets them exchange information through
multi-head self-attention and then acts on each stream individually with a multilayer perceptron.
The output streams $\{\boldsymbol{h}^L_i\}$ are finally projected to produce the amplitudes $F^n_j$ and the winding maps $\eta^{(\alpha,n)}_j$. We follow
\cite{abouelkomsan2026first} for the explicit forms with hyperparameters listed in Table~\ref{tab:hyperparameters}.

We improve the expressivity of the network by replacing the purely permutation-symmetric winding
maps of \cite{abouelkomsan2026first} by
\begin{equation}
    \eta^{(\alpha,n)}_j(\mathbf{r}_i;\{\mathbf{r}_{\neq i}\}) =
    u^{(\alpha,n)}_{2j}\!\cdot\!\boldsymbol{h}^L_i
        + i\,u^{(\alpha,n)}_{2j+1}\!\cdot\!\boldsymbol{h}^L_i
    + \frac{1}{N-1}\sum_{k \neq i}
        \left[v^{(\alpha,n)}_{2j}\!\cdot\!\boldsymbol{h}^L_k
            + i\,v^{(\alpha,n)}_{2j+1}\!\cdot\!\boldsymbol{h}^L_k\right],
    \label{eq:eta_rowdep}
\end{equation}
with $u,v$ two independent matrices. The first term uses the stream of particle $i$ and the second one uses the mean over the remaining streams. This allows a richer $\mathbf{r}_i$ dependence of the winding maps without breaking the equivariance of the orbitals $\phi^n_j$. The twist constraint $\sum_\alpha \eta^{(\alpha,n)}_j = 0$ is imposed as before, by subtracting the $\alpha$ mean. 

\begin{table*}[h]
\caption{Hyperparameters used in this work.}
\label{tab:hyperparameters}
\centering
\begin{tabular}{l@{\hskip 1em}r@{\hskip 3em}l@{\hskip 1em}r@{\hskip 3em}l@{\hskip 1em}r}
\toprule
\multicolumn{2}{l}{\textit{Architecture}} &
\multicolumn{2}{l}{\textit{Optimization}} &
\multicolumn{2}{l}{\textit{Monte Carlo}} \\
\cmidrule(r{3em}){1-2}\cmidrule(r{3em}){3-4}\cmidrule{5-6}
Layers $L$          & 4    & Optimizer     & MinSR    & Walkers               & 1024 \\
Heads               & 4    & Learning rate & 0.02     & Steps/sample          & 10   \\
Attention dim.      & 32   & Iterations    & $10^3$ to $10^4$ & Move width  & $\ell_B$ \\
Perceptron dim.     & 128  &               &          & Burn-in               & 300  \\
Perceptrons/layer   & 2    &               &          &                       &      \\
Determinants        & 2    &               &          &                       &      \\
Activation          & GELU &               &          &                       &      \\
\bottomrule
\end{tabular}
\end{table*}

\section{Landau-level occupations}\label{sec:LLocc}

Landau-level mixing is what distinguishes the network wave function from a calculation restricted to a few levels. To measure it, we expand the one-body density matrix of the trained state $\Phi_{\mathbf{k}}$ in the Landau-level orbitals $\varphi_{n,\mathbf{q}}$ of Sec.~\ref{sec:orbitals}. For a normalized $\Psi$ of $N$ electrons,
\begin{equation}
\begin{split}
  \rho_{(n,\mathbf{q}),(n',\mathbf{q}')} = N \int &d\mathbf{r}_1\, d\mathbf{r}'_1\, d\mathbf{r}_2 \cdots d\mathbf{r}_{N} \; \overline{\varphi_{n,\mathbf{q}}(\mathbf{r}_1)}\, \Psi(\mathbf{r}_1, \mathbf{r}_2, \ldots)\; \varphi_{n',\mathbf{q}'}(\mathbf{r}'_1)\, \overline{\Psi(\mathbf{r}'_1, \mathbf{r}_2, \ldots)} .
\end{split}
\label{eq:rdm_def}
\end{equation}
The configurations $\{\mathbf{r}_i\}$ are sampled from $|\Psi|^2$, and the $\mathbf{r}'_1$ integral is done on a uniform grid over the supercell. We take the Hermitian part $\tfrac12(\rho + \rho^\dagger)$ to reduce the noise. The number of electrons in level $n$ and its filling are
\begin{equation}
  N_n = \sum_{\mathbf{q}} \rho_{(n,\mathbf{q}),(n,\mathbf{q})}, \qquad \nu_n = N_n/N_\phi ,
\end{equation}
compared with $\nu^{(0)} = (1,\tfrac12,0,\dots)$, the filling without Landau-level mixing. $N_n$ is a trace over the level and so does not depend on the choice of basis within it.

\section{Trial states and Schrieffer--Wolff dressing}
\label{sec:dressed}

To identify the topological order of the NN-VMC ground-state manifold, we compute its overlaps with trial states for the Pfaffian, anti-Pfaffian, and PH-Pfaffian orders, as well as the ground states of the 1LL-projected Coulomb interaction. The latter provide a particle--hole symmetric reference. While these trial states are defined in a Hilbert space with a completely filled LLL and the remaining electrons restricted to the 1LL, the NN-VMC states include Landau level mixing. To account for this difference, we also consider trial states dressed by a Schrieffer--Wolff transformation, which incorporates virtual excitations between Landau levels.

\subsection{Model space}

We define the model space $\mathcal{P}$ by completely filling the LLL and restricting the remaining electrons to the 1LL. A basis for this space is given by the Slater determinants
\begin{equation}
    D_I(\mathbf{r}_1, \ldots, \mathbf{r}_{N}) = \frac{1}{\sqrt{N!}}\,\det\big[\varphi_{0,\mathbf{q}_1}, \ldots, \varphi_{0,\mathbf{q}_{N_\phi}}, \varphi_{1,\mathbf{p}_1}, \ldots, \varphi_{1,\mathbf{p}_{\tilde{N}}}\big](\mathbf{r}_1, \ldots, \mathbf{r}_{N}),
    \label{eq:DI}
\end{equation}
where $\varphi_{n,\mathbf{q}}$ are the Landau-level orbitals defined in Sec.~\ref{sec:orbitals}, and $I = \{\mathbf{p}_1, \ldots, \mathbf{p}_{\tilde{N}}\}$ specifies the occupied 1LL orbitals of the remaining $\tilde{N} = N - N_\phi$ electrons.

We write each trial state as $\ket{\mathrm{X}} = \sum_I c^{\mathrm{X}}_I \ket{D_I}$. The coefficients $c^{\mathrm{X}}_I$ are obtained from the corresponding state of $\tilde{N}$ electrons in a half-filled Landau level and are used to construct the state in the 1LL, with the filled LLL treated as an inert background. For the system sizes considered here, all trial manifolds have the same dimension and occupy the same momentum sectors as the NN-VMC ground-state manifold [Fig.~\ref{fig:entropy}(a)].

\subsection{Pfaffian and anti-Pfaffian}

We construct the Pfaffian manifold from the zero modes of the three-body parent Hamiltonian~\cite{greiter1991paired, rezayi2000incompressible}
\begin{equation}
    H_3 = -\sum_{i<j<k} \mathcal{S}_{ijk}\Big[\nabla_i^4\,\nabla_j^2\,\delta^2(\mathbf{r}_i - \mathbf{r}_j)\,\delta^2(\mathbf{r}_j - \mathbf{r}_k)\Big],
    \label{eq:H3}
\end{equation}
where $\mathcal{S}_{ijk}$ denotes symmetrization over the three particles. We diagonalize this Hamiltonian for $\tilde{N}$ electrons in $N_\phi$ orbitals using LLL form factors, which gives the Pfaffian zero modes.

The anti-Pfaffian is obtained by particle--hole conjugation of the Pfaffian within the half-filled Landau level~\cite{levin2007particle, lee2007particle}. We define the antiunitary particle--hole transformation $\mathcal{C}$ through $\mathcal{C}\,c_{\mathbf{q}}\,\mathcal{C}^{-1} = c^\dagger_{\mathbf{q}}$. Acting on a Slater determinant, it exchanges occupied and empty orbitals, $\mathcal{C}\ket{I} = s(I)\ket{\bar{I}}$, where $s(I)=\pm 1$ accounts for the fermionic sign, while also complex conjugating the coefficients. The anti-Pfaffian manifold is therefore obtained from $\mathcal{C}\ket{\mathrm{Pf}}$, or equivalently as the zero modes of $\mathcal{C}H_3\mathcal{C}^{-1}$. In practice, we apply particle--hole conjugation to the full many-body matrix of $H_3$, including the one-body and two-body terms generated by the transformation.

\subsection{PH-Pfaffian}

For the PH-Pfaffian~\cite{son2015composite}, we start from the wave function $\mathrm{Pf}[1/(\bar{z}_i-\bar{z}_j)]\prod_{i<j}(z_i-z_j)^2$~\cite{zucker2016stabilization, mishmash2018numerical}. On the torus, we construct the corresponding unprojected states using the dimensionless coordinates $\zeta=z/\sqrt{2}\ell_B$,
\begin{equation}
    \Psi^{\mathrm{PH}}_{a,k} = \overline{\mathrm{Pf}\big[g_a(\zeta_i - \zeta_j)\big]}\;\prod_{i<j}\sigma(\zeta_i - \zeta_j)^2\;\sigma\big(Z - W_{a,k}/2\big)^2\;\prod_i e^{-|\zeta_i|^2/2 - \frac{N_\phi}{2}\mathcal{A}\,\zeta_i^2 + b_{a,k}\,\zeta_i},
    \label{eq:phpf}
\end{equation}
where $\sigma$ is the Weierstrass sigma function of the supercell and $Z=\sum_i\zeta_i$. The pair function is given by $g_a(\zeta)=\vartheta_a(u|\tau)/\vartheta_1(u|\tau)$, where $\vartheta_a$ are the Jacobi theta functions, $u=\pi\zeta/2w_1$, $\tau=w_2/w_1$, and $w_{1,2}=\mathbf{L}_{1,2}/2\sqrt{2}\ell_B$ are the complex half-periods of the torus. The index $a=2,3,4$ labels the three even spin structures, while $k$ labels the two independent center-of-mass factors for each spin structure. The parameters $\mathcal{A}$, $b_{a,k}$, and $W_{a,k}$ are chosen to satisfy the magnetic boundary conditions in the appropriate gauge.

Without complex conjugation of the Pfaffian factor, Eq.~\eqref{eq:phpf} reduces to the Moore--Read state on the torus~\cite{read2000paired}. The complex conjugation does not affect the magnetic boundary conditions, since the quasiperiodicity multipliers of $g_a$ are real. We combine the resulting six states into momentum eigenstates.

Because of the antiholomorphic Pfaffian factor, Eq.~\eqref{eq:phpf} is not confined to a single Landau level. Unlike the Pfaffian and the anti-Pfaffian, it therefore has no unique counterpart in the half-filled 1LL, and we consider three constructions.

\begingroup\renewcommand{\thesubsubsection}{\upshape\arabic{subsubsection}}%
\subsubsection{\texorpdfstring{\upshape LLL-projected, raised to the 1LL}{LLL-projected, raised to the 1LL}}
We project Eq.~\eqref{eq:phpf} onto the LLL by computing its overlaps with all $\tilde{N}$-electron LLL Slater determinants $D_I^{(0)}$. These overlaps are evaluated by Monte Carlo sampling from $|\Psi^{\mathrm{PH}}|^2$, using
\begin{equation}
    \left\langle
    \overline{\frac{D_I^{(0)}(x)}{\Psi^{\mathrm{PH}}(x)}}
    \right\rangle_{|\Psi^{\mathrm{PH}}|^2}
    =
    \frac{\langle D_I^{(0)}|\Psi^{\mathrm{PH}}\rangle}
    {\langle\Psi^{\mathrm{PH}}|\Psi^{\mathrm{PH}}\rangle}.
\end{equation}
This procedure projects the full many-body wave function, rather than projecting its individual factors separately. The resulting expansion coefficients are then used to construct the corresponding states in the 1LL. In the LLL, this projected wave function was found to be numerically very similar to the composite Fermi liquid~\cite{mishmash2018numerical}.

\subsubsection{\texorpdfstring{\upshape Unprojected, raised to the 1LL}{Unprojected, raised to the 1LL}}
We raise every electron of Eq.~\eqref{eq:phpf} by one Landau level, $n \to n+1$, keeping the expansion coefficients, and place the result on top of the filled LLL. The resulting state extends over several Landau levels. We truncate it to the lowest few and verify that the overlap is stable as the truncation is relaxed. Since this state already has components outside $\mathcal{P}$, we dress it by applying $e^{(\lambda/\kappa)S}$ of Eq.~\eqref{eq:dressing} to the full state. The components outside $\mathcal{P}$ that lie in the span of the excitations $TD_I$ are then rotated back into $\mathcal{P}$, while the others are unchanged.

\subsubsection{\texorpdfstring{\upshape Unprojected}{Unprojected}}
We antisymmetrize Eq.~\eqref{eq:phpf} directly with the filled LLL. The Pauli principle removes all components with an electron in the LLL, and the component of the resulting state in $\mathcal{P}$ is the second-Landau-level projection of the PH-Pfaffian~\cite{yang2023second}. We use this component, which can be dressed in the same way as the other trial states. It is not particle--hole symmetric and describes pairing in the $l = -3$ channel, the pairing channel of the anti-Pfaffian~\cite{yang2023second, zhao2023composite}.
\endgroup

\subsection{Coulomb states}

As a particle--hole symmetric reference, we use the low-energy manifold obtained by exact diagonalization of the Coulomb interaction projected to the 1LL. Within $\mathcal{P}$, the filled LLL contributes only a constant to the projected Hamiltonian. These states are therefore also the lowest-energy eigenstates of Eq.~\eqref{eq:hamiltonian} restricted to $\mathcal{P}$.

\subsection{Schrieffer--Wolff dressing}

We now incorporate Landau level mixing into the trial states through a Schrieffer--Wolff transformation. In units of $\hbar\omega_c$, the Hamiltonian in Eq.~\eqref{eq:hamiltonian} can be written as $H=H_0+\kappa V$, where $H_0=\sum_i(a_i^\dagger a_i+\tfrac12)$ is the kinetic energy and
\begin{equation}
    V = \frac{1}{2A}\sum_{\mathbf{q}\neq 0} \frac{2\pi}{|\mathbf{q}|}\,{:}\rho(-\mathbf{q})\,\rho(\mathbf{q}){:}
    \label{eq:Vcoulomb}
\end{equation}
is the normal-ordered Coulomb interaction in units of $U$, with lengths measured in $\ell_B$. Here $A$ is the area of the supercell and $\mathbf{q}$ runs over its reciprocal lattice vectors, with the $\mathbf{q}=0$ contribution cancelled by the neutralizing background. Writing the electron position as $\mathbf{r} = \mathbf{R} + \tilde{\mathbf{R}}$, with the commuting guiding-center coordinate $\mathbf{R}$ and cyclotron coordinate $\tilde{\mathbf{R}}$, the density operator in the Landau-level basis is
$\rho(\mathbf{q})=\sum_{nn'}\sum_{\mathbf{k}\mathbf{k}'}F_{nn'}(\mathbf{q})\,\bar{R}_{\mathbf{k}\mathbf{k}'}(\mathbf{q})\,c^\dagger_{n\mathbf{k}}c_{n'\mathbf{k}'}$.
Here $\bar{R}_{\mathbf{k}\mathbf{k}'}(\mathbf{q}) = \langle\mathbf{k}|e^{i\mathbf{q}\cdot\mathbf{R}}|\mathbf{k}'\rangle$ is the guiding-center translation matrix between the momentum states $\mathbf{k}$ and $\mathbf{k}'$, and $F_{nn'}(\mathbf{q}) = \langle n|e^{i\mathbf{q}\cdot\tilde{\mathbf{R}}}|n'\rangle$ are the Landau-level form factors,
\begin{equation}
    F_{nn'}(\mathbf{q}) = \sqrt{\frac{n'!}{n!}}\left(\frac{i q_+}{\sqrt{2}}\right)^{n-n'} e^{-|\mathbf{q}|^2/4}\, L^{(n-n')}_{n'}\!\left(\frac{|\mathbf{q}|^2}{2}\right),
    \label{eq:formfactor}
\end{equation}
for $n\geq n'$, where $L^{(\alpha)}_{n}$ are the generalized Laguerre polynomials, $q_+=q_x+iq_y$, and $F_{n'n}(\mathbf{q})=\overline{F_{nn'}(-\mathbf{q})}$.

Since all states in $\mathcal{P}$ have the same kinetic energy $E_0$, their coupling to states outside $\mathcal{P}$ arises entirely from the interaction. Introducing the projectors $P$ onto $\mathcal{P}$ and $Q=1-P$, we construct a unitary transformation $e^S$ that eliminates the off-diagonal couplings $PVQ$ and $QVP$ to first order in $\kappa$~\cite{schrieffer1966relation, bravyi2011schrieffer}. The transformed Hamiltonian $e^{-S}He^S$ is then block diagonal to this order, while the corresponding states in the full Hilbert space are obtained by acting with $e^S$ on states in $\mathcal{P}$.

The generator satisfies $[H_0,S]=-\kappa(PVQ+QVP)$, giving
\begin{equation}
    S = -\kappa\,\big(T-T^\dagger\big), \qquad
    T=Q\,(H_0-E_0)^{-1}\,Q\,V\,P .
    \label{eq:SWgenerator}
\end{equation}
The operator $T$ contains all interaction-induced pair scatterings from $\mathcal{P}$ into states with different Landau-level occupations. These processes involve changes in the Landau-level index of one or two electrons, with a net increase of at least one cyclotron quantum. They include, for example, excitations from the filled LLL that leave behind a hole, as well as excitations of electrons from the 1LL to higher Landau levels. The energy denominators are determined by the change in cyclotron energy, without including interaction corrections.

To dress a trial state $\ket{\mathrm{X}}$, we exponentiate the first-order generator, allowing its strength to vary through a parameter $\lambda$,
\begin{equation}
    \ket{\tilde{\mathrm{X}}}
    =e^{(\lambda/\kappa)S}\ket{\mathrm{X}}
    =\cos\!\big(\lambda\sqrt{\Gamma}\big)\ket{\mathrm{X}}
    -T\,\Gamma^{-1/2}\sin\!\big(\lambda\sqrt{\Gamma}\big)\ket{\mathrm{X}},
    \qquad
    \Gamma=T^\dagger T
    =P\,V\,Q\,(H_0-E_0)^{-2}\,Q\,V\,P .
    \label{eq:dressing}
\end{equation}
The expression follows from the off-diagonal structure of $T$, which maps $\mathcal{P}$ to its complement, while $T^\dagger$ maps these excitations back into $\mathcal{P}$. Since the transformation is unitary, the dressed states remain normalized.

For $\lambda=\kappa$, expanding Eq.~\eqref{eq:dressing} to first order gives
\begin{equation}
    \ket{\tilde{\mathrm{X}}}
    =
    \ket{\mathrm{X}}
    -
    \kappa Q(H_0-E_0)^{-1}QV\ket{\mathrm{X}}
    +O(\kappa^2).
\end{equation}
For the Coulomb states, which are eigenstates of $PVP$, this reproduces the first-order perturbative correction to the wave function outside $\mathcal{P}$. For the other trial states, the dressing incorporates the same virtual inter-Landau-level processes.

We evaluate $\Gamma$ by explicitly enumerating all determinants coupled to $\mathcal{P}$ through $V$, including Landau levels up to a cutoff $n_c$. We verify convergence with respect to this cutoff by monitoring the weight of the virtual excitations, $\kappa^2\langle\mathrm{X}|\Gamma|\mathrm{X}\rangle$, which is converged to within a few percent.

For each trial state and value of $\kappa$, we determine $\lambda$ by maximizing the overlap with the NN-VMC ground-state manifold. This gives a one-parameter variational dressing of the trial states. We find that the perturbative choice $\lambda=\kappa$ overestimates the Landau level mixing at large $\kappa$, resulting in substantially smaller overlaps. The optimal $\lambda/\kappa$ decreases as $\kappa$ increases and takes nearly the same value for the Pfaffian, anti-Pfaffian, and Coulomb states. For the LLL-projected PH-Pfaffian, the overlaps remain small, making the optimal $\lambda$ poorly constrained at small $\kappa$.

\subsection{Overlaps}

We compare the NN-VMC ground-state manifold with each trial manifold through the overlap defined in the main text, $\mathcal{O} = d^{-1}\sum_{\alpha,\beta=1}^{d}|\langle\Phi_\alpha|\mathrm{X}_\beta\rangle|^2$, where $\{\ket{\Phi_\alpha}\}$ and $\{\ket{\mathrm{X}_\beta}\}$ are orthonormal bases of the $d$-dimensional NN-VMC and trial manifolds. We evaluate it both for the bare trial states and for the dressed trial states, for which $\ket{\mathrm{X}_\beta}$ is replaced by $\ket{\tilde{\mathrm{X}}_\beta}$. Since states in different momentum sectors are orthogonal, $\mathcal{O}$ is a sum of contributions from the individual sectors $\mathbf{k}$. The states we compute within a sector are not exactly orthonormal, and we therefore write $\mathcal{O}$ in terms of the Gram matrices $G_\Phi$ and $G_{\mathrm{X}}$ of the two sets,
\begin{equation}
    \mathcal{O}
    =\frac{1}{d}\sum_{\mathbf{k}}
    \mathrm{Tr}\big[G_\Phi^{-1}M\big],
    \qquad
    M_{ij}
    =
    \sum_{t,t'}
    \langle\Phi_i|\mathrm{X}_t\rangle\,
    \big(G_{\mathrm{X}}^{-1}\big)_{tt'}\,
    \langle\mathrm{X}_{t'}|\Phi_j\rangle,
    \label{eq:overlapgram}
\end{equation}
which reduces to the expression above when both sets are orthonormal. The Pfaffian, anti-Pfaffian and Coulomb states are eigenvectors of Hermitian matrices and are therefore orthonormal, so that $G_{\mathrm{X}}$ is the identity. The PH-Pfaffian states, obtained by projection or truncation, are neither normalized nor mutually orthogonal, and $G_{\mathrm{X}}$ accounts for their overlaps. Since the dressing is unitary, the dressed trial states have the same Gram matrix as the bare ones.

The matrix $M$ requires the overlaps of each NN-VMC state $\Phi$ with the trial states, both bare and dressed. For a bare trial state, the overlap follows from the projections of $\Phi$ onto the determinants of $\mathcal{P}$, $\langle\Phi|\mathrm{X}\rangle = \sum_I c^{\mathrm{X}}_I\,\langle\Phi|D_I\rangle$. For a dressed trial state, Eq.~\eqref{eq:dressing} also involves the projections of $\Phi$ onto the interaction-induced excitations $\chi_I=TD_I$ of these determinants,
\begin{equation}
    \langle\Phi|\tilde{\mathrm{X}}\rangle
    =\sum_I\Big[
    \langle\Phi|D_I\rangle\,
    \big(\cos(\lambda\sqrt{\Gamma})\,c^{\mathrm{X}}\big)_I
    -
    \langle\Phi|\chi_I\rangle\,
    \big(\Gamma^{-1/2}\sin(\lambda\sqrt{\Gamma})\,c^{\mathrm{X}}\big)_I
    \Big],
    \label{eq:dressedoverlap}
\end{equation}
which reduces to the bare overlap for $\lambda = 0$. Both sets of amplitudes, $\langle\Phi|D_I\rangle$ and $\langle\Phi|\chi_I\rangle$, are estimated by Monte Carlo sampling from $|\Phi|^2$.

\subsection{Overlaps with the PH-Pfaffian}
\begin{figure}[h]
  \centering
  \includegraphics[width=0.6\textwidth]{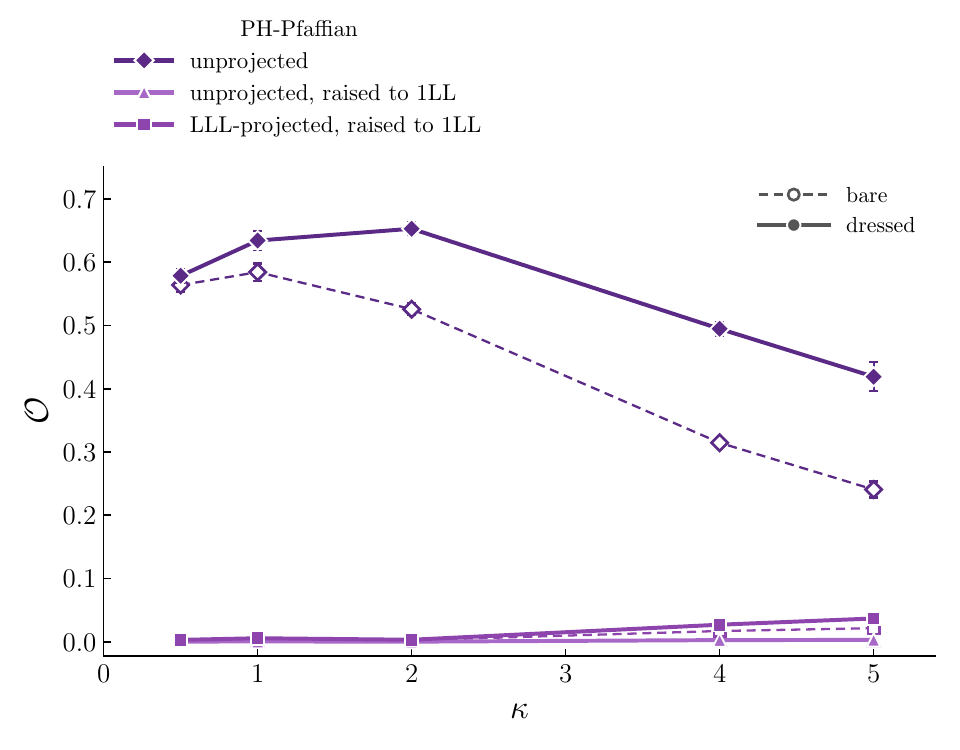}
  \caption{Overlap $\mathcal{O}$ between the NN-VMC ground-state manifold of $N = 18$ electrons and the three constructions of the PH-Pfaffian described in Sec.~\ref{sec:dressed}, as a function of Landau level mixing $\kappa$. Open symbols with dashed lines show bare overlaps, and filled symbols with solid lines show dressed overlaps.}
  \label{fig:phpf}
\end{figure}

Figure~\ref{fig:phpf} shows the overlaps of the three PH-Pfaffian constructions with the NN-VMC ground-state manifold. The LLL-projected and the raised unprojected constructions have negligible overlaps at all $\kappa$, both bare and dressed. The unprojected construction has a sizable overlap, smaller than that of the anti-Pfaffian in Fig.~\ref{fig:pfapf} at small $\kappa$ and comparable to it at large $\kappa$, consistent with its anti-Pfaffian pairing channel.

\section{Fine magnetic translations within a momentum sector}\label{sec:penalty}

The momentum projection of Sec.~\ref{sec:bc} uses only the translations generated by $\mathbf{a}_1$ and $\mathbf{a}_2$. The remaining translations on the fine lattice $\Lambda_\phi$, the grey points of Fig.~\ref{fig:tilings}, also commute with $H$ and preserve the boundary conditions. They carry additional quantum numbers that further organize the spectrum within each sector $\mathbf{k}$. We now describe these quantum numbers and how we use them to resolve more states and improve convergence.

\subsection{Commuting translations on the fine lattice.}
By the composition rule of Sec.~\ref{sec:bc}, the center-of-mass translations on $\Lambda_\phi$ obey
\begin{equation}
  T(\mathbf{a})\,T(\mathbf{b}) = e^{\,2\pi i N f_{\mathbf{a}\mathbf{b}}}\, T(\mathbf{b})\,T(\mathbf{a}), \qquad f_{\mathbf{a}\mathbf{b}} = \frac{\hat{z}\cdot(\mathbf{a}\times\mathbf{b})}{2\pi\ell_B^{2}},
  \label{eq:comtranslationalgebra}
\end{equation}
where $f_{\mathbf{a}\mathbf{b}}$ is the number of flux quanta through the cell spanned by $\mathbf{a}$ and $\mathbf{b}$. A cell of $\Lambda_\phi$ carries $f = 1/N_\phi$, and a sublattice of $\Lambda_\phi$ whose cell contains $s$ of its points carries $f = s/N_\phi$. For $\nu = N/N_\phi = p/q$ with $p$ and $q$ coprime, $N f = p\,s/q$, so all translations of the sublattice commute when $q$ divides $s$. The largest commuting sets are therefore the minimizing $s$ to $s = q$. They contain $N_\phi^2/q = N_0 N_\phi$ translations, with $N_0 = \gcd(N, N_\phi) = N_\phi/q$. 

The translations of Sec.~\ref{sec:bc} generated by $\mathbf{a}_1$ and $\mathbf{a}_2$ belong to a maximal set generated by $T(\mathbf{a}_1/n_1)$ and $T(\mathbf{a}_2/n_2)$ with $n_1 n_2 = N_0$, shown in blue in Fig.~\ref{fig:tilings}. Since $T(\mathbf{a}_j/n_j)^{n_j} = T(\mathbf{a}_j)$, the eigenvalues of these generators are the $n_j$-th roots of the eigenvalues $\lambda_j(\mathbf{k})$ of Eq.~\eqref{eq:momenta}, $\mu_{l_j} = \lambda_j(\mathbf{k})^{1/n_j}\, e^{2\pi i l_j/n_j}$ with $l_j = 0, \dots, n_j - 1$. The integers $(l_1, l_2)$ are the additional quantum numbers, and each sector $\mathbf{k}$ splits into $N_0 = n_1 n_2$ fine sectors, with $(n_1, n_2) = (3, 2)$ for $N = 18$ on the $4 \times 3$ tiling and $(1, 7)$ for $N = 21$ on the $7 \times 2$ tiling.

\subsection{Fine-momentum projection.}
\begin{figure*}
  \centering
  \includegraphics[width=\textwidth]{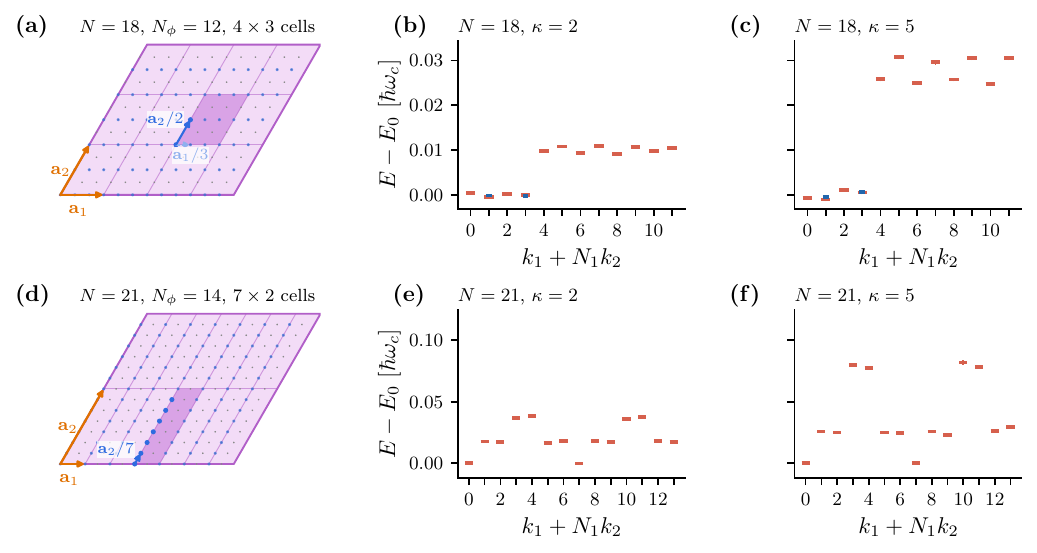}
  \caption{Supercell tilings and fine-projected spectra, one row per system. (a) The tiling of $N=18$ ($4\times3$
  cells) and (d) of $N=21$ ($7\times2$ cells). The one-flux cell vectors $\mathbf{a}_1$, $\mathbf{a}_2$ are in
  orange, the fine lattice $\Lambda_\phi$ in grey, and the maximal commuting set of translations in blue. In one cell
  the generator $T(\mathbf{a}_2/n_2)$ of the fine projection is drawn with its $n_2$ points in dark blue. The unused
  $\mathbf{a}_1/3$ of $N=18$ is dashed. (b, c) NN-VMC spectra of $N=18$ at $\kappa=2$ and $\kappa=5$. Each bar is the
  energy of the fine-projected state $\Phi_{\mathbf{k},l}$ of Eq.~\eqref{eq:fineprojection} in one momentum sector,
  with its statistical error. Energies are relative to the mean $E_0$ of the sixfold ground multiplet,
  $E_0 = -4.9736 \pm 0.0001\,\hbar\omega_c$ (b) and $-37.1050 \pm 0.0002\,\hbar\omega_c$ (c). The sectors $(1,0)$ and $(3,0)$ hold two
  states: the second is drawn in blue, shorter, over the first. (e, f) The same for $N=21$ at $\kappa=2$ and
  $\kappa=5$. Here $E_0$ is the mean of the ground doublet ($k_1+N_1k_2=0,7$), $E_0 = -6.5028 \pm 0.0003\,\hbar\omega_c$ (e)
  and $-45.0552 \pm 0.0003\,\hbar\omega_c$ (f).}
  \label{fig:tilings}
\end{figure*}

Projecting onto a full maximal commuting set would fix every quantum number at once, but would multiply the cost by $N_0^2$, since each amplitude would need $N_0 N_\phi$ network evaluations instead of $N_\phi$, and there would be $N_0 N_\phi$ fine sectors to optimize instead of $N_\phi$. We therefore resolve a single additional quantum number, the eigenvalue $\mu_l$ of the generator $T(\mathbf{a}_2/n_2)$ (solid arrow in Fig.~\ref{fig:tilings}), with $l = l_2$, through the projection
\begin{equation}
  \Phi_{\mathbf{k},l} = \frac{1}{n_2} \sum_{j=0}^{n_2-1} \mu_l^{-j}\, T(\mathbf{a}_2/n_2)^{\,j}\, \Phi_{\mathbf{k}}.
  \label{eq:fineprojection}
\end{equation}
Each $\Phi_{\mathbf{k},l}$ is an exact eigenstate of $T(\mathbf{a}_2/n_2)$ and states with different $l$ are orthogonal by construction. For $N = 21$ this resolves all translation quantum numbers, since $n_1 = 1$. For $N = 18$, $\mathbf{a}_2/2$ is the finest translation along $\mathbf{a}_2$ that commutes with $T(\mathbf{a}_1)$, and the label $l_1$ of $T(\mathbf{a}_1/3)$ is left unresolved.

\subsection{Training.}
In each of the $N_\phi$ sectors we first optimize $\Phi_{\mathbf{k}}$ with the sector projection alone. In general, a single optimization converges to an uncontrolled mixture of several $l$ components. The weight of each component $\Phi_{\mathbf{k},l}$ is its overlap with the optimized state $\Phi_{\mathbf{k}}$. To obtain the spectra of Fig.~\ref{fig:spectrum} and Fig.~\ref{fig:entropy}, we select the $l$ carrying the largest weight and continue the optimization with the fine-projected ansatz $\Phi_{\mathbf{k},l}$ of Eq.~\eqref{eq:fineprojection} at that value of $l$. Optimizing $\Phi_{\mathbf{k},l}$ at the remaining values of $l$ yields further members of a multiplet, orthogonal to the first by construction. These are the partners (blue) that complete the ground-state manifold of $N = 18$ in Fig.~\ref{fig:entropy}.

\subsection{Orthogonality penalty.}
In a given sector, two states can share every translation quantum number. The $N = 21$ crystal of Sec.~\ref{sec:crystals} is such a case: its ground state and its first excited state lie in the same fine sector, and the fine projection cannot separate them. We therefore also use an orthogonality penalty to resolve multiplets sequentially. The state $\Phi_a$ under optimization is penalized for its overlap with the fixed partners $\{\Phi_b\}$ through the modified loss
\begin{equation}
  \mathcal{L} = E_{\mathbf{k}}(\theta) + \lambda \sum_{b} |S_{ab}|^2, \qquad S_{ab} = \frac{\langle \Phi_a | \Phi_b \rangle}{\|\Phi_a\|\,\|\Phi_b\|},
\end{equation}
which for large enough $\lambda$ drives the minimum orthogonal to the states already resolved~\cite{pathak2021excited}. Unlike the fine projection of Eq.~\eqref{eq:fineprojection}, this route does not fix $l$: it targets the lowest state orthogonal to those already found regardless of its $l$ content. 

\section{Wigner crystals at $N = 21$}\label{sec:crystals}

At strong Landau-level mixing the electrons are expected to depart from the liquid state and form a Wigner crystal~\cite{price1993fractional,zhao2018crystallization,abouelkomsan2026first}. A triangular lattice of $N$ sites fits the hexagonal torus only if $N = a^2 + ab + b^2$ for integers $a, b$. No such lattice exists for $N = 18$. For $N = 21 = 4^2 + 4 + 1$ there are two consistent lattices, mirror images of each other  rotated by $+10.9^\circ$ and $-10.9^\circ$ from $\mathbf{L}_1$, which we call $A$ and $B$ [Fig.~\ref{fig:decat}(a) and (b)]. $N = 21$ thus hosts both phases without frustration: the ground state is a liquid at $\kappa = 5$ and a crystal at $\kappa = 20$. This section presents the $\kappa = 20$ spectrum and explains it with the fine translations of Sec.~\ref{sec:penalty}.

\subsection{Fine sectors.}
For $N = 21$ on the $7 \times 2$ tiling of Fig.~\ref{fig:tilings}, $N_\phi = 14$, $N_0 = 7$, and the maximal commuting set is generated by $T(\mathbf{a}_1)$ and $T(\mathbf{a}_2/7)$. We write the $\Phi_{\mathbf{k}}$ sectors as $\mathbf{k} = (k_1, k_2)$ with $k_1 = 0, \dots, 6$ and $k_2 = 0, 1$, so that $\lambda_1(\mathbf{k}) = e^{2\pi i k_1/7}$ and $\lambda_2(\mathbf{k}) = e^{i\pi k_2}$. The fine-projected states $\Phi_{\mathbf{k},l}$ of Sec.~\ref{sec:penalty} are then the eigenstates of the generators with 
\begin{equation}
T(\mathbf{a}_1)\,\Phi_{\mathbf{k},l} = \lambda_1(\mathbf{k})\,\Phi_{\mathbf{k},l}, \qquad
T(\mathbf{a}_2/7)\,\Phi_{\mathbf{k},l} = \mu_l\,\Phi_{\mathbf{k},l}, \qquad
  \mu_l = \lambda_2(\mathbf{k})^{1/7}\, e^{2\pi i l/7} = e^{2\pi i (k_2 + 2l)/14}, \quad l = 0, \dots, 6 .
  \label{eq:finesectors21}
\end{equation}
For each $k_2$ there are $49$ fine sectors $(k_1, l)$, which form the two grids of Fig.~\ref{fig:decat}(c).

\subsection{Crystal sectors, Fig.~\ref{fig:decat}(c).}
Consider a rigid triangular crystal $\Psi_A$ with the periodicity of lattice $A$. Its translation symmetries differ in general from the fine translations of the tiling [Fig.~\ref{fig:tilings}(d)] and a crystal is in general not an eigenstate of the ansatz's translations. However, there are fine translations that are lattice vectors of the crystal making special fine sectors natural hosts for crystal states. Within the maximal commuting set for $N = 21$ these translations are powers of $T_A = T(\mathbf{a}_1)\,T(\mathbf{a}_2/7)^4$, which translates every electron by $\mathbf{a}_1 + 4\mathbf{a}_2/7$ [arrow in Fig.~\ref{fig:decat}(a)].

A fine sector $(\mathbf{k}, l)$ can host the crystal only if the projection $P_{\mathbf{k},l}\Psi_A$ onto that sector is nonzero. The projector can be written explicitly as 
\begin{equation}
  P_{\mathbf{k},l} = \frac{1}{98} \sum_{p=0}^{6} \sum_{q=0}^{13} \lambda_1(\mathbf{k})^{-p}\, \mu_l^{-q}\; T(\mathbf{a}_1)^p\, T(\mathbf{a}_2/7)^q.
  \label{eq:projector}
\end{equation}
The action of $T_A$ on the projection is expressed with Eq.~\eqref{eq:finesectors21} as 
\begin{equation}
  T_A\, [P_{\mathbf{k},l}\Psi_A] = \tau\, [P_{\mathbf{k},l}\Psi_A], \qquad \tau = \lambda_1(\mathbf{k})\,\mu_l^{\,4}.
  \label{eq:tau}
\end{equation}
Since $T_A$ commutes with every term in $P_{\mathbf{k},l}$ and $T_A \Psi_A = \Psi_A$, we have
\begin{equation}
  P_{\mathbf{k},l}\,\Psi_A = P_{\mathbf{k},l}\, T_A\, \Psi_A = T_A\, P_{\mathbf{k},l}\,\Psi_A = \tau\, P_{\mathbf{k},l}\,\Psi_A
  \label{eq:crystalproj}
\end{equation}
which imposes the projection vanishes unless $\tau = 1$. It follows that a rigid crystal of orientation $A$ exists only in the fine sectors where $T_A$ has eigenvalue $1$, and there the projected state keeps the symmetry of the crystal. For $B$ the same holds with $T_B = T(\mathbf{a}_1)\,T(\mathbf{a}_2/7)^8$, which translates by $\mathbf{a}_1 + 8\mathbf{a}_2/7$. With Eq.~\eqref{eq:finesectors21} the two crystal compatibility conditions read
\begin{equation}
  A:\ k_1 + 2k_2 + 4l \equiv 0, \qquad B:\ k_1 + 4k_2 + l \equiv 0 \pmod 7 .
  \label{eq:crystallines}
\end{equation}
Fig.~\ref{fig:decat}(c) draws the two conditions on the grids of fine sectors: the red line is $A$, and the blue line is $B$. Both wrap around the grid because the labels are periodic modulo $7$. Each line passes through seven fine sectors, and the two lines intersect once for each $k_2$, at $k_1 = 0$ and $l = 3k_2$, the black point, which we call the origin. Away from the origin, a point on a line allows one orientation and excludes the mirror one by its quantum numbers alone. At the origin both orientations are allowed, and they can mix. Off the lines no rigid crystal of either orientation exists.

\subsection{Spectrum, Fig.~\ref{fig:decat}(d).}
In the liquid phase, at $\kappa = 1$, $2$ and $5$, the lowest excitations lie off the crystal lines, and every state has the pair correlations of a liquid. The $\kappa = 20$ spectrum is different. Figure~\ref{fig:decat}(d) shows the lowest fine-projected state on the $A$ line and on the $B$ line in each sector.

Twelve of the fourteen sectors do not hold the ground-state manifold; we call them the excited sectors, with $k_1 \neq 0$. In each excited sector the two lowest fine-projected states are a crystal of orientation $A$ and a crystal of orientation $B$. They carry different fine quantum numbers $l$, one on each line of Fig.~\ref{fig:decat}(c), and they are all degenerate, at $+0.14\,\hbar\omega_c$ above the ground state.

A flat manifold spanned by the translated and mirrored copies of one ordered state is a finite-size signature of spontaneous symmetry breaking~\cite{yang2001wigner}. The crystal breaks the fine translations, down to its own lattice translations, and the mirror that exchanges $A$ and $B$. We note that in a magnetic field this mirror is an anti-unitary combination of reflection with time reversal.

An exception to this symmetry-breaking pattern occurs at the origin of Fig.~\ref{fig:decat}(c), corresponding to the ground-state sectors of Fig.~\ref{fig:decat}(d). There, the $A$ and $B$ lines cross and both orientations share all quantum numbers allowing them to hybridize. The resulting ground state $G$ lies $0.14\,\hbar\omega_c$ below the band, and a second lowest state $P$, obtained with the penalty method of Sec.~\ref{sec:penalty} against $G$, lies $0.11\,\hbar\omega_c$ above it. The pair correlations of $G$ and $P$ shown in Fig.~\ref{fig:decat}(e) reveal the hybridization of the $A$ and $B$ orientations. We now show that this pair of states is an Anderson tower~\cite{anderson1952approximate}.

Let $\Psi_A$ and $\Psi_B$ be the rigid crystals of the two orientations projected onto the origin, Eq.~\eqref{eq:crystalproj} at $k_1 = 0$, $l = 3k_2$. They share all quantum numbers, so $H$ is free to couple them through $t_{AB} = \langle \Psi_A | H | \Psi_B \rangle$ and the eigenstates in the space they span are superpositions $a\Psi_A + b\Psi_B$. The mirror commutes with $H$ and exchanges $\Psi_A$ and $\Psi_B$ imposing $|a| = |b|$. The eigenstates are then equal-weight superpositions $\Psi_A \pm \Psi_B$ (cat states) of the two orientations, split by $2|t_{AB}|$.

This hybridization is a finite-size effect. The two orientations are macroscopically distinct configurations of the $N$ electrons, so their overlap and the tunneling amplitude $t_{AB}$ both fall exponentially with $N$. The splitting at the origin is thus expected to close in the thermodynamic limit.

\begin{figure*}
  \centering
  \includegraphics[width=\textwidth]{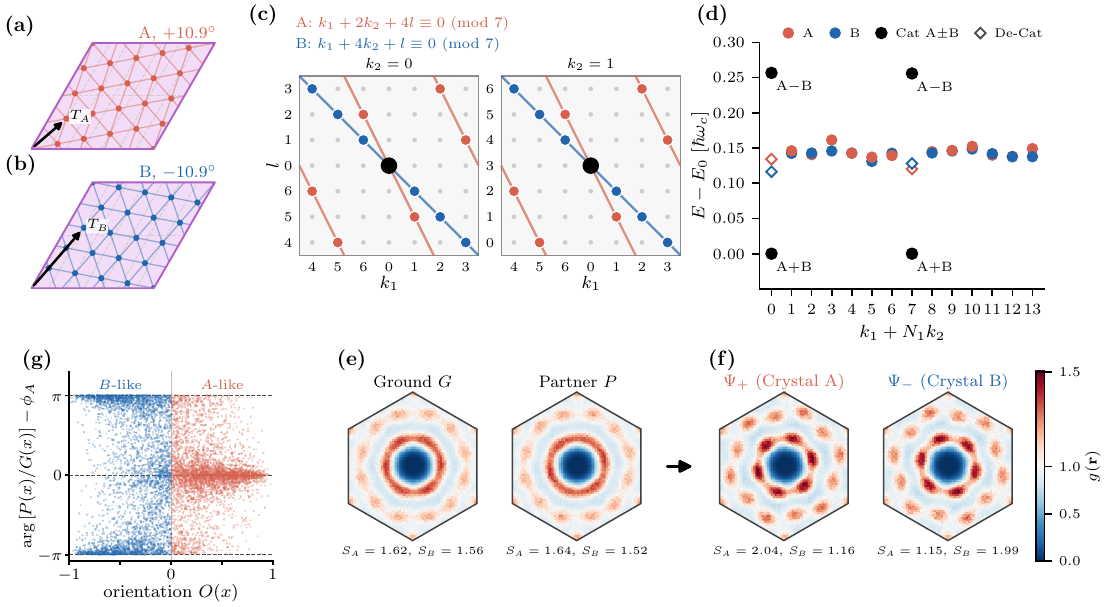}
  \caption{Wigner crystals at $N = 21$, $\kappa = 20$. (a, b) The two $21$-site triangular lattices $A$ and $B$ of the $7 \times 2$ torus, rotated by $\pm 10.9^\circ$ from $\mathbf{L}_1$, and the fine translation that preserves each of them: $T_A = T(\mathbf{a}_1)\,T(\mathbf{a}_2/7)^4$, by $\mathbf{a}_1 + 4\mathbf{a}_2/7$, and $T_B = T(\mathbf{a}_1)\,T(\mathbf{a}_2/7)^8$, by $\mathbf{a}_1 + 8\mathbf{a}_2/7$ (arrows). (c) The fine sectors $(k_1, l)$ of each sector $k_2 = 0, 1$, both labels modulo $7$. A rigid crystal of orientation $A$ ($B$) can exist only on the red (blue) line, Eq.~\eqref{eq:crystallines}; the lines cross only at the origin $k_1 = 0$, $l = 3k_2$ (black). (d) Spectrum relative to $E_0 = -252.819\,\hbar\omega_c$, the mean energy of the two ground states $G$: in each sector the lowest fine-projected state on the $A$ line (red) and on the $B$ line (blue); at the origin, columns $0$ and $7$, the cat pair, the ground state $G$ (A$+$B) and its penalty partner $P$ (A$-$B), in black, and the energies of the De-Cat states $\Psi_\pm = (G \pm e^{-i\phi_A} P)/\sqrt{2}$ (open diamonds). (g) The sign flip in the sector $k_2 = 0$: the phase of $P(x)/G(x)$, measured from its value on the $A$ side, against the orientation $O(x) = [S_A(x) - S_B(x)]/[S_A(x) + S_B(x)]$ of the sampled configurations $x$, with $S_X$ of Eq.~\eqref{eq:orientation}. (e) Pair correlation $g(\mathbf{r})$ and Bragg weights $S_A$, $S_B$ of $G$ and $P$: both superpose the two orientations. (f) The same as (e) for $\Psi_+$ and $\Psi_-$: single crystals $A$ and $B$.}
  \label{fig:decat}
\end{figure*}

\subsection{De-Cat, Fig.~\ref{fig:decat}(e,f).}
We now test this picture on the two states themselves, by separating their superposed $A, B$ orientations. $G$ is the fine-projected ground state at the origin and $P$ its penalty partner at the same quantum numbers, Sec.~\ref{sec:penalty}. 

To see the two orientations combined in $G$ and $P$ we look at the individual configurations $x = (\mathbf{r}_1, \dots, \mathbf{r}_N)$ of the Monte Carlo sampling drawn from $|G|^2$ and from $|P|^2$. Let $Q_X$ be the set of the six shortest reciprocal-lattice vectors of the lattice $X = A, B$. The structure factor of a configuration on $Q_X$,
\begin{equation}
  S_X(x) = \frac{1}{6N} \sum_{\mathbf{q} \in Q_X} \Big| \sum_{j=1}^{N} e^{i\mathbf{q}\cdot\mathbf{r}_j} \Big|^2 ,
  \label{eq:orientation}
\end{equation}
measures the order of orientation $X$ in that one configuration, and its average over the configurations of a state is the Bragg weight $S_X$ of that state. A single crystal has $S_X = 2.0$ for its own orientation and $1.2$ for the other. $G$ and $P$ both have $S_A \simeq S_B \simeq 1.6$, and their pair correlations in Fig.~\ref{fig:decat}(e) show the two orientations superposed.

If $G$ and $P$ are the cat states of $\Psi_A$ and $\Psi_B$, we expect $G = (\Psi_A + \Psi_B)/\sqrt{2}$ and $P = e^{i\phi_A}(\Psi_A - \Psi_B)/\sqrt{2}$, with a relative phase $\phi_A$ that the optimization leaves arbitrary. The ratio $\rho(x) = P(x)/G(x)$ then equals $e^{i\phi_A}$ on the configurations where $\Psi_A$ dominates and $-e^{i\phi_A}$ where $\Psi_B$ dominates. To take advantage of this property, we sort the configurations by their orientation $O(x) = [S_A(x) - S_B(x)]/[S_A(x) + S_B(x)]$, which is positive for $A$-like configurations and negative for $B$-like ones. Figure~\ref{fig:decat}(g) shows the phase of $\rho$ against $O(x)$.
It is constant for $O > 0.5$, constant for $O < -0.5$, and the two values differ by $0.98\pi$ in both sectors $k_2$.

This reveals the cat-state structure, and it gives reconstructions of $\Psi_A$ and $\Psi_B$ as
\begin{equation}
  \Psi_\pm = \frac{G \pm e^{-i\phi_A} P}{\sqrt{2}} .
  \label{eq:rotation}
\end{equation}
Indeed, the pair correlations of $\Psi_\pm$ in Fig.~\ref{fig:decat}(f) are those of single crystals, $(S_A, S_B) = (2.04, 1.16)$ and $(1.15, 1.99)$, and their energies, the open diamonds of Fig.~\ref{fig:decat}(d), lie at $+0.12$ to $+0.13\,\hbar\omega_c$, inside the crystal band, where the pair is expected to end up as $N$ grows. The pair correlation of $\Psi_+$ is the one shown in Fig.~\ref{fig:LLmixing}(b) of the main text, where it makes the crystalline order at $\kappa = 20$ manifest. The $\kappa = 20$ ground state is the symmetric resonance of the two orientations, its partner is the antisymmetric one, and every crystal that cannot resonate sits $|t_{AB}| \simeq 0.13\,\hbar\omega_c$ above it.

\end{document}